\documentclass[floatfix,pra,twocolumn,aps,superscriptaddress,showpacs]{revtex4-2}
\usepackage{graphicx}
\usepackage[normalem]{ulem}
\usepackage{mathrsfs,amsmath}
\usepackage{amsfonts}
\usepackage{amssymb}
\usepackage{epsfig}
\usepackage{subfigure}
\usepackage{mathtools}
\usepackage{braket}
\usepackage[usenames,dvipsnames]{color}
\usepackage{setspace}%\input{response_to_referees}
\usepackage{bm}
\usepackage{times}
\usepackage{ulem}
\usepackage{dsfont}
\usepackage{xcolor}
\usepackage{adjustbox}
\usepackage{hyperref}
\hypersetup{
	colorlinks=true,
	citecolor=blue,
	linkcolor=blue,
	urlcolor=blue}
    
\begin{document}
\title[] {Spin and density excitations of one-dimensional self-bound Bose-Bose droplets}
% \author{Ritu, Manpreet Singh, Rajat, Rajesh Kumar Gupta, and Sandeep Gautam}
% \affiliation{Department of Physics, Indian Institute of Technology Ropar, Rupnagar 140001, Punjab, India}
\author{Ritu}\email{ritu.22phz0002@iitrpr.ac.in}
\affiliation{Department of Physics, Indian Institute of Technology Ropar, Rupnagar 140001, Punjab, India}
\author{Rajat}\email{rajat13sep@gmail.com}
\affiliation{Department of Physics, Indian Institute of Technology Ropar, Rupnagar 140001, Punjab, India}
\affiliation{Department of Physics, Ben-Gurion University of the Negev, Beer-Sheva 84105, Israel} 
\author{Manpreet Singh}\email{manpreet.25phz0008@iitrpr.ac.in}
\affiliation{Department of Physics, Indian Institute of Technology Ropar, Rupnagar 140001, Punjab, India}
\author{Rajesh Kumar Gupta}\email{rajesh.gupta@iitrpr.ac.in}
\affiliation{Department of Physics, Indian Institute of Technology Ropar, Rupnagar 140001, Punjab, India}
\author{Sandeep Gautam}\email{sandeep@iitrpr.ac.in}
\affiliation{Department of Physics, Indian Institute of Technology Ropar, Rupnagar 140001, Punjab, India}
\begin{abstract}
We study spin and density excitations of one-dimensional self-bound Bose-Bose droplets within Bogoliubov theory, and show that spin excitations come alive, especially
as the interspecies coupling is made less attractive. We argue that spin excitations are particularly relevant in the one-dimensional droplet regime, where droplets are realized within the mean-field stability regime, as 
has been confirmed by the quantum Monte Carlo simulations. As the interspecies coupling strength increases within the mean-field stability regime, 
spin modes ultimately fall below the particle-emission threshold, thus becoming observable in the droplet spectrum. 
We analyze the Bogoliubov model for both pseudospinor and population-imbalanced scalar mixtures, encompassing both the density and spin sectors. We corroborate our findings through variational analysis of density- and spin-breathing modes, which offers physical insight into the mode structure and independently validates the spectrum, as well as through real-time dynamics. Additionally, we compare our results with both Petrov's original theory, which considers the Lee-Huang-Yang (LHY) correction at the attractive edge of the mean-field stability regime, and a beyond-LHY description of Bose-Bose mixtures.
\end{abstract}

\maketitle
\section{Introduction}
\label{intro}
Quantum droplets are a self-bound phase of ultracold quantum gases that arises from beyond-mean-field effects. This phase was first predicted by Petrov~\cite{PhysRevLett.115.155302} and was later experimentally realized in homonuclear mixtures of two components~\cite {doi:10.1126/science.aao5686, PhysRevLett.120.235301, PhysRevLett.120.135301, PhysRevLett.122.090401} and heteronuclear mixtures~\cite{PhysRevResearch.1.033155, Wang_PhysRevResearch.3.033247, Burchianti_PhysRevLett.134.093401}, as well as in dipolar Bose-Einstein condensates ~\cite{Kadau2016, PhysRevLett.116.215301, Schmitt2016, PhysRevX.6.041039}. Droplets are stabilized when 
the mean-field interactions and Lee-Huang-Yang (LHY) correction provide competing contributions to the energy density, preventing
the mean-field collapse in these systems \cite{PhysRevLett.115.155302, petrov_notes}. 

In three dimensions, the LHY term is repulsive, so droplets require net attractive mean-field interactions, i.e.,
$\delta g = \sqrt{g_{11} g_{22}}+g_{12} <0,~|\delta g|\ll \sqrt{g_{11}g_{22}}=g$, where $\left(g_{11},g_{22}\right)$ and $g_{12}$ are the intra- and interspecies coupling constants, respectively \cite{PhysRevLett.115.155302}. The LHY energy can be written as a sum over zero-point energies of Bogoliubov modes; however, for $\delta g<0$, the Bogoliubov dispersion reveals a dynamically unstable density (in-phase) branch and a stable spin (out-of-phase) branch
~\cite{pethick_smith_2008,pitaevskii2016bose}. Including both branches naively yields an unphysical complex LHY energy~\cite{PhysRevLett.115.155302, 10.21468/SciPostPhys.9.2.020}. Petrov's prescription avoids this by discarding the density-branch contribution 
and retaining only the stable branch when constructing the LHY energy~\cite{PhysRevLett.115.155302}. 
The mean-field energy density reveals that the mean intraspecies coupling constant $(g_{11}+g_{22})/2$ sets the stiffness of the spin fluctuations, 
while much smaller $\delta g \sqrt{g_{11} g_{22}}/(g_{11}+g_{22})$ governs the stiffness of the density fluctuations. Since $|\delta g|/g \ll 1$ for a three-dimensional (3D) droplet, this hierarchy of stiffnesses leads to a separation of energy scales between density and spin channels, pushing the spin excitations above the particle-emission threshold \cite{PhysRevLett.115.155302,petrov_notes,   PhysRevA.102.053303}. 
These considerations permit an effective single-component description of the homonuclear droplets
with frozen density ratio. Later theoretical studies explored the two-component nature of droplets beyond the single-mode approximation, considering, for instance, LHY correction for the heteronuclear 3D droplets at the mean-field collapse threshold~\cite{PhysRevA.98.053623, PhysRevA.100.063636}, stability analysis beyond the density-locking approximation~\cite{PhysRevA.103.013312}, 
and the role of population imbalance on the ground states and density-breathing modes in mass-balanced 3D droplets~\cite{PhysRevResearch.5.033167, PhysRevResearch.6.013209}. Motivated by Ref.~\cite{PhysRevLett.120.135301}, a study of the collective modes across the soliton-droplet crossover in a quasi-one-dimensional waveguide has shown that particle emission inhibits the modes over a wide range of particle numbers and interactions in the droplet regime, with the density-breathing mode as the only accessible mode over a small particle range and intermediate interactions \cite{PhysRevA.97.053623}. A generalization of Bogoliubov theory with the inclusion of bosonic pairing \cite{Hui_PhysRevLett.125.195302, PhysRevA.102.043301}, and beyond Bogoliubov theories \cite{10.21468/SciPostPhys.9.2.020, PhysRevB.102.220503} have also been proposed to remedy the issue of the imaginary speed for the density mode. 

Quantum fluctuations also lead to the droplet phase in 2D Bose-Bose mixtures with weak interspecies attraction and weak intraspecies repulsion \cite{PhysRevLett.117.100401, petrov_notes}.
The 2D quantum droplets hosting vortices \cite{Malomed_PhysRevA.98.063602},  vortex clusters under rotating confinements \cite{Reimann_PhysRevLett.123.160405, Oktel_PhysRevA.108.033315}, and a variety of other non-linear excitations \cite{Mistakidis_PhysRevA.110.033317}  have been theoretically studied in addition to studies of modulation instability \cite{PhysRevA.106.033309} and excitation spectrum \cite{PhysRevA.103.053302, PhysRevA.106.053303, PhysRevA.109.053309}.

In contrast to three and two dimensions, 1D droplets are stabilized by the attractive LHY correction balancing the net mean-field repulsion ($\delta g>0$)~\cite{PhysRevLett.117.100401}.
The droplets are therefore realized within the mean-field stability regime with a stable density phonon mode, removing any need to ignore its contribution to the LHY correction~\cite{10.21468/SciPostPhys.9.2.020}. 
Most of the studies on 1D droplets still consider an approximate LHY correction evaluated at $g_{12} = -g$
as per Petrov's original prescription, 
which makes the corresponding description suitable for $0<\delta g\ll g$. At $\delta g \gtrsim 0$, the spin excitations lie above the particle emission threshold, allowing a single-component description of 1D droplets \cite{PhysRevLett.117.100401, PhysRevA.101.051601, PhysRevA.98.013631}. The single-component description of 1D droplets has been used to study ground states ~\cite{PhysRevLett.117.100401}, the excitation spectrum ~\cite{PhysRevA.101.051601, PhysRevA.98.013631, PhysRevA.108.033312}, collision dynamics~\cite{PhysRevA.98.013631}, and solitary waves \cite{Edmonds_PhysRevResearch.5.023175, Kevrekidis_PhysRevA.107.063308}.
As $\delta g$ increases within the mean-field stability regime, spin excitations soften and may lie below the particle emission threshold. This avenue is constrained for
3D droplets, where $\delta g$ is negative and close to the mean-field collapse threshold, thus maintaining the separation of scales between the ``soft" density and ``stiff" spin channels.

In the regime with accessible density and spin excitations, one has to treat the droplet as a genuine two-component system.
The two-component description, albeit with an approximate LHY correction (Petrov's prescription), has been used to study the rotational properties of scalar Bose-Bose mixtures confined on a ring with a particle imbalance \cite{Reimann_PhysRevA.105.033319}.
The description has also been used to study the stability of 1D droplets under weak harmonic confinement in scalar mixtures with a particle imbalance \cite{PhysRevA.111.013318}.
The LHY corrections for a weakly interacting single-component Bose gas and Bose-Bose mixtures across the crossover from three to two dimensions and from three dimensions to one have also been calculated \cite{PhysRevA.98.051604, PhysRevA.98.051603, PhysRevA.103.033312}. 

The predictions of the Bogoliubov theory %\sout{(and Petrov's model)} 
deviate significantly from quantum Monte Carlo (QMC) results for both 3D and 1D droplets, with the discrepancy increasing as $\delta g$ moves away from zero~\cite{PhysRevA.99.023618, PhysRevLett.122.105302, PhysRevA.102.023318}. Moreover, in one dimension, Bogoliubov theory predicts self-bound droplets throughout the entire mean-field-stable regime. This is in stark contrast to QMC calculations, which instead predict a liquid-to-gas transition when $\delta g$ exceeds a critical value~\cite{PhysRevLett.122.105302, PhysRevA.102.023318}. This transition has attracted considerable recent interest~\cite{PhysRevLett.130.193001, PhysRevA.97.063616, PhysRevA.103.043316}. A more accurate, beyond-LHY description of self-bound droplets is therefore desirable. Such a description, proposed for pseudospinor mixtures in Ref.~\cite{10.21468/SciPostPhys.9.2.020}, self-consistently incorporates higher-order corrections to the sound velocities, providing very good agreement with the QMC results. The present work provides an extension of this beyond-LHY description to a scalar mixture with population imbalance. 
The impact of beyond-LHY correlations on the ground state and the dynamics of 1D droplets has been studied using an \textit{ab initio}
nonperturbative approach \cite{Schmelcher_PhysRevResearch.3.043128,Schmelcher_PhysRevA.107.023320,PhysRevA.110.023324}. 

Recently, Ref.~\cite{10.21468/SciPostPhys.19.5.133} discussed the comparison between the ground state
density profiles obtained with Bogoliubov theory and Petrov's original prescription across the entire mean-field stability regime. These theories
match at $\delta g = 0$, and as $\delta g$ is increased, the latter significantly overestimates the finite-size droplets' saturation densities, confining them over a smaller extent.
In this work we use the Bogoliubov theory with exact LHY correction to calculate the density and spin excitations of 1D Bose-Bose droplets. 
The explicit behavior of the spin excitation spectrum in self-bound, lower-dimensional geometries has yet to be clarified. In this work we %build on these foundational frameworks by showing 
show that the 1D geometry %, combined with the exact LHY correction, 
 provides a unique setting in which out-of-phase spin modes can be continuously tuned below the particle-emission threshold by varying the interspecies interaction.
We calculate the full excitation spectrum from
the numerical solutions of Bogoliubov equations and analyze the lowest-lying density- and spin-breathing modes using a variational analysis. We consider both self-bound pseudospinor and 
population-imbalanced scalar Bose-Bose mixtures. 
The results of Bogoliubov theory are compared with those from Petrov's model and, in addition, for the density-breathing mode, with the beyond-LHY theory. 

The paper is organized as follows. We introduce the Bogoliubov theory used for the description of 1D Bose-Bose mixtures and contrast it with Petrov's original prescription in Sec. \ref{Sec-I}. We discuss pseudospinor Bose-Bose mixtures, revisiting a simpler description, which ignores the spin excitations, 
and later their excitations are obtained numerically using a full two-component description and supported by a variational analysis
in Sec. \ref{Sec-II}. This is followed by a discussion of the excitations in a scalar Bose-Bose mixture in Sec. \ref{Sec-III}.
We then discuss a beyond-LHY description of 1D Bose-Bose mixtures to examine the liquid-to-gas transition and the density breathing mode
in Sec. \ref{Sec-IV}. We summarize this work in Sec. \ref{summary}.

\section{Model}
\label{Sec-I}
We consider a 1D two-component weakly interacting Bose-Bose mixture within the framework described
by the Bogoliubov approximation \cite{PhysRevLett.117.100401, petrov_notes}. 
The energy density of such a homogeneous mixture 
with equal masses of the two species is \cite{PhysRevLett.117.100401}
\begin{equation}
    {\cal E} =\frac{1}{2}\sum_{\nu \nu'}g_{\nu \nu'}n_\nu n_{\nu'} - \frac{2\sqrt{m}}{3\hbar\pi}\sum_{\pm}c_{\pm}^3,
    \label{ener}
\end{equation}
where $\nu$ and $\nu'$ are the labels (1 or 2) of the two components, $g_{\nu\nu'}$ denotes the coupling constants, $n_\nu$
is the density of the species $\nu$, and 
$c_{\pm}$ are the two sound velocities. We consider the system in the weak coupling limit, for which the dimensionless parameter $n|a_{\nu\nu'}|\gg1$, where $n=\sum_\nu n_\nu$ and
$a_{\nu\nu'}$ are $s$-wave scattering lengths related with coupling constants as $g_{\nu\nu'} = -2\hbar^2/ma_{\nu\nu'}$ \cite{PhysRevLett.117.100401, PhysRevLett.122.105302}.
Experimental realization of such 1D mixtures would require tighter confinement along the transverse directions to ensure that transverse trapping energy, $\hbar\omega_{\perp}$, 
is much larger than the other relevant energy scales in the system, such as $k_BT$ and $\mu$. 
The weak coupling limit can then potentially be achieved  if $a_{\perp}/a_{\nu\nu'}^{3{\rm D}}$ is sufficiently large, where $a_{\perp}$ is the confinement length scale along
the transverse direction and $a_{\nu\nu'}^{3{\rm D}}$ are 3D $s$-wave scattering lengths \cite{pitaevskii2016bose}.
In this context, it is worthwhile to point out that the Tonks-Girardeau gas, where $n|a_{\nu\nu'}|\ll1$, has already been experimentally realized \cite{science.1100700, Paredes2004}.

In Eq. (\ref{ener}), the first term is the mean-field contribution to the energy density and the second is the LHY
correction arising from the quantum fluctuations with $c_{\pm}^2$ given as 
\begin{align}
    c_{\pm}^2=\frac{g_{11}n_1+g_{22}n_2\pm \sqrt{(g_{11}n_1-g_{22}n_2)^2+4g_{12}^2n_1n_2}}{2}.
\end{align}
The finite-size self-bound solutions (solitons or droplets) of the system can be described by the Lagrangian
\begin{align}
L =& \int dx \Bigg[\sum_\nu \left\{\frac{i\hbar}{2}\left(\Phi_\nu^* \frac{\partial \Phi_\nu}{\partial t}-\Phi_\nu\frac{\partial \Phi_\nu^*}{\partial t}\right)-\frac{\hbar^2}{2m}\left|\frac{d \Phi_\nu}{dx}\right|^2\right\} \nonumber\\
&-{\cal E}\Bigg],
\label{Lag}
\end{align}
where $\Phi_\nu(x,t)$ are the component wavefunctions with $|\Phi_\nu(x,t)|^2 = n_\nu(x,t)$. The Euler-Lagrange or Gross-Pitaevskii (GP) equations describing the dynamics of the system are \cite{10.21468/SciPostPhys.19.5.133} 
\begin{equation}
    i\hbar\frac{\partial\Phi_\nu}{\partial t} =\Bigg( -\frac{\hbar^2}{2m}\frac{\partial^2}{\partial x^2} + g_{\nu\nu} |\Phi_\nu|^2 + g_{\nu \bar{\nu}} |\Phi_{\bar{\nu}}|^2 + \Delta\mu_\nu\Bigg) \Phi_\nu,
    \label{eeGPE}
\end{equation}
where $\bar \nu = 1, 2$; $\bar \nu\ne \nu$ and
\begin{align}
    \Delta\mu_\nu &= -\frac{\sqrt{m}}{\pi\hbar}\sum_{\pm}c_\pm\frac{\partial c_\pm^2}{\partial n_\nu},\label{mu_lhy}\\&= -\frac{\sqrt{m(g_{11}n_1 + g_{22}n_2)}}{4\pi\hbar} \nonumber\\ 
    &\quad \times \bigg[ 3 g_{\nu\nu} \mathcal{R}(p) + 2 (g_{11}n_1 + g_{22}n_2) \frac{\partial \mathcal{R}}{\partial p} \frac{\partial p}{\partial n_\nu} \bigg],\label{lhy}
\end{align}
%\end{equation}
is the LHY correction term.
In Eq.~(\ref{lhy}), $p$ and $\mathcal{R}(p)$ are defined as
\begin{subequations}
    \begin{align}
    \label{lhy_var}
    p&=\frac{4(g_{12}^2-g_{11}g_{22})n_1n_2}{(g_{11}n_1+g_{22}n_2)^2},\\
        \mathcal{R}(p) &= \frac{\sqrt{2}}{3}\Big((1-\sqrt{p+1})^\frac{3}{2}+(1+\sqrt{p+1})^\frac{3}{2}\Big). \label{Rp}       
    \end{align}
\end{subequations}
If ${\cal R}(p)$ in Eq.~(\ref{Rp}) is evaluated at the attractive edge of the mean-field stability regime, i.e., $g_{12}=-\sqrt{g_{11}g_{22}}$ or $p=0$,

Equation~(\ref{eeGPE}) reduces to 
\begin{equation}
\label{oeGPE}
\begin{split}
    i\hbar\frac{\partial\Phi_\nu}{\partial t} &= \biggl( -\frac{\hbar^2}{2m}\frac{\partial^2}{\partial x^2} + g_{\nu \nu}|\Phi_\nu|^2 + g_{\nu\bar{\nu}} |\Phi_{\bar{\nu}}|^2 \\
    &\quad - \frac{\sqrt{m}g_{\nu\nu}}{ \hbar\pi}\sqrt{g_{11}|\Phi_1|^2+g_{22}|\Phi_2|^2} \biggr)\Phi_\nu.
\end{split}
\end{equation}
This is the Euler-Lagrange equation corresponding to the Lagrangian (\ref{Lag}) with $\cal E$ now denoting the energy density 
obtained by Petrov and Astrakharchik in Ref.~\cite{PhysRevLett.117.100401}. 
We refer to Eqs.~(\ref{eeGPE}) and (\ref{oeGPE}) as Bogoliubov's and Petrov's models, respectively. Under the single-mode approximation, Eq.~(\ref{oeGPE}) reduces to the single-component equation derived in Ref.~\cite{PhysRevLett.117.100401} [our Eq.~(\ref{dl_gpe})]. It should be noted that the LHY correction within Bogoliubov theory 
depends on $g_{12}$, unlike the model used by Petrov. It modifies the saturation density and the excitation spectrum, as shown below.
In this work we consider pseudospinor and scalar Bose-Bose mixtures.
If $N_\nu$ denotes the number of particles of species $\nu$ and $N = \sum_\nu N_\nu$ is the total number of particles in 
a finite-size droplet, then 
\begin{equation}
\int \sum_\nu |\Phi_\nu(x,t)|^2 dx = N,
\end{equation}
for a pseudospinor mixture, whereas
\begin{equation}
\int |\Phi_\nu(x,t)|^2 dx = N_\nu,
\end{equation}
for a scalar mixture. 
To calculate the excitation spectrum, we linearize Eq.~(\ref{eeGPE}) by considering fluctuation 
around the ground state $\phi_\nu(x)$ using
\begin{equation}
    \Phi_\nu(x,t) = e^{-i \mu_\nu t}\Big[\phi_\nu(x) + \delta{\phi}_{\nu}(x,t)\Big],
\end{equation}
where $\mu_{\nu}$ are the chemical potentials; $\mu_1 = \mu_2 = \mu$ for a pseudospinor mixture.
Then, using the Bogoliubov transformation 
\begin{align}
    \delta{\phi}_{\nu}(x,t)=\sum_{l}\left[u_\nu^l (x) e^{-i \omega_l t}-v_\nu^{l*}(x)e^{i\omega_l t} \right] \label{fluc},
\end{align}
we obtain the Bogoliubov–de Gennes (BdG) equations as
\\
\\
\begin{widetext}
\begin{equation}\label{bdg}
\begin{split}
\begin{pmatrix}
 \mathcal{M} & -g_{11}\phi_1^2 - B_{11} & g_{12}\phi_1\phi_2^* + A_{21} & -g_{12}\phi_1\phi_2 - B_{21} \\
 g_{11}\phi_1^2 + B_{11}^* & -\mathcal{M}^* & g_{12}\phi_1^*\phi_2^* + B_{21}^* & -g_{12}\phi_1^*\phi_2 - A_{21}^* \\ 
 g_{21}\phi_2\phi_1^* + A_{12} & -g_{21}\phi_2\phi_1 - B_{12} & \mathcal{N} & -g_{22}\phi_2^2 - B_{22} \\ 
 g_{21}\phi_2^*\phi_1^*+B_{12}^* & -g_{21}\phi_2^*\phi_1-A_{12}^* & g_{22}\phi_2^{*2}+B_{22}^* & -\mathcal{N}^*
\end{pmatrix}
\begin{pmatrix}
 u_1^l \\ v_1^l \\ u_2^l \\ v_2^l
\end{pmatrix}
&= \omega_l
\begin{pmatrix}
 u_1^l \\ v_1^l \\ u_2^l \\ v_2^l
\end{pmatrix},
\end{split}
\end{equation}
\end{widetext}
where elements $\mathcal{M}$, $\mathcal{N}$, $A$, and $B$ for the two models are defined by Eqs.~(\ref{matrix_element}), (\ref{matrix_element_petrov}) and (\ref{matrix_element_bogo}) in Appendix A.
The quasiparticle wave functions with positive norm satisfy the normalization condition \cite{pethick_smith_2008}
\begin{equation} \label{bdg_norm}
\sum_{\nu}\int dx\left[|u_\nu^l(x)|^2 -|v_\nu^{l}(x)|^2\right] = 1,
\end{equation}

where $l$ represents the eigenvalue index corresponding to energy $\omega_l$ with $u_\nu^{l}$ and $v_\nu^{l}$ as quasi-particle wave functions.
We return to the limits of validity of the Bogoliubov treatment and to the liquid-to-gas transition in Sec. \ref{Sec-IV}.

\section{Two-component pseudospinor mixture}
\label{Sec-II}
{\em Bogoliubov theory within single-mode approximation}. In a 1D pseudospinor mixture near the attractive edge of the mean-field stability regime, where 
$0<\delta g = g_{12}+\sqrt{g_{11}g_{22}}\ll \sqrt{g_{11}g_{22}}$, the spin excitations are energetically costlier relative to the density excitations
and may well lie above the particle-emission threshold of approximately $|\delta g| n$ for $\delta g$ close to zero. 
In this regime, the ground-state structure and low-lying density excitations can be studied within a simpler framework provided by the single-mode approximation (SMA) \cite{PhysRevLett.115.155302, PhysRevLett.117.100401, PhysRevA.101.051601, PhysRevA.98.013631}. 
The SMA entails that the component wavefunctions are proportional to a single wave function, i.e., $\Phi_\nu(x,t)= \alpha_\nu \Phi(x,t)$, where
$|\Phi|^2 = \sum_\nu n_\nu = n$ and $\alpha_\nu$ are $c$ numbers with $\sum_{\nu} |\alpha_\nu|^2 = 1$. The constant density ratio
$n_1/n_2 = |\alpha_1|^2/(1-|\alpha_1|^2) = \beta$ within the SMA. To calculate $\beta$, we rewrite the energy density of the homogeneous 
mixture within the Bogoliubov theory as
\begin{subequations}
\begin{align}
    \mathcal{E} =& \frac{(g_{11}n_1^2+g_{22}n_2^2)}{2}+g_{12}n_1n_2
                -\frac{\sqrt{m}(g_{11}n_1+g_{22}n_2)^\frac{3}{2}}{2\hbar\pi}\nonumber\\&\times\mathcal{R}(p),\\
    =& \frac{n^2 (g_{11}\beta^2+g_{22}+2g_{12}\beta)}{2(\beta+1)^2}
    -\frac{\sqrt{m(g_{11}\beta+g_{22})^3}n^\frac{3}{2}}{2\hbar\pi\sqrt{(\beta +1)^3}}\nonumber\\&\times\mathcal{R}(p),
    \label{enden}
\end{align}
\end{subequations} 
%\sout{where $\zeta =4\beta(g_{12}^2-g_{11}g_{22})/{(g_{11}\beta+g_{22})^2}$,}
 where $p$ in Eq.~(\ref{lhy_var}) is expressed in terms of $\beta$, whereas the mean-field energy density within Petrov's model corresponds
 to $p = 0$ or $\mathcal {R} (p = 0) = 4/3$ in Eq.~(\ref{enden}). 
The minimization of energy per particle ${\cal E}(n,\beta)/n$ with respect to 
$(n,\beta)$ for fixed $g_{11}, g_{22},$ and $ g_{12}$ yields the saturation density of the flat-top droplets and the precise density ratio. 
In Refs. \cite{PhysRevLett.115.155302,PhysRevLett.117.100401, PhysRevA.101.051601, PhysRevA.98.013631}, the density-locking condition has been 
approximated as $n_1/n_2 = \sqrt{g_{22}/g_{11}}$, which holds true at $\delta g= 0$ or trivially for $g_{11} = g_{22}$, to arrive at an effective single-component description.
\begin{figure}[!htbp]
\centering
\includegraphics[width=\columnwidth]{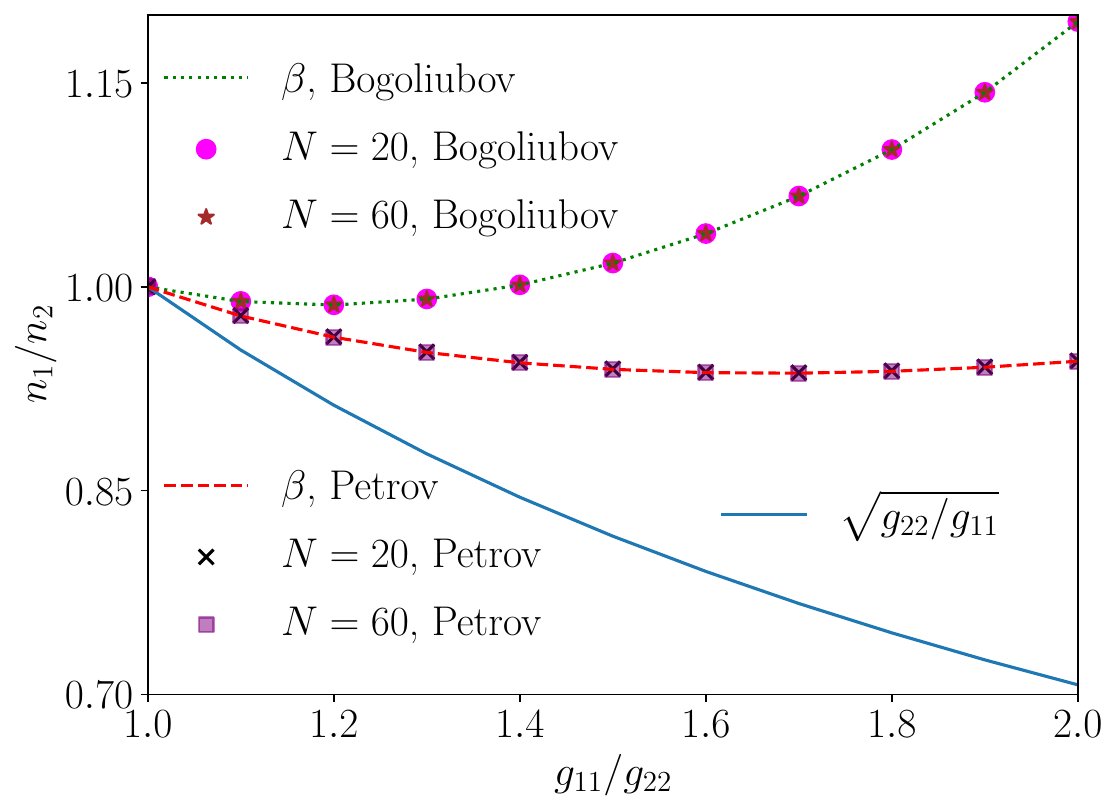}
\caption{Density ratio $\beta = n_1/n_2$ obtained by a minimization of the energy per particle $\mathcal{E}(n,\beta)/n$ as a function  $g_{11}/g_{22}$ for $g_{12} = -0.6 g_{22}$
for the Bogoliubov and Petrov theories. Symbols denote the density ratio at the center of the finite-size self-trapped solutions with $N = 20$ and $60$ obtained
by numerically solving the GP equations~(\ref{eeGPE}) for the Bogoliubov and (\ref{oeGPE}) for the Petrov theory. For comparison, the simpler estimate of the density ratio $\sqrt{g_{22}/g_{11}}$,
which minimizes the mean-field energy density for $g_{12} = -\sqrt{g_{11}g_{22}}$ is also shown.
}
\label{beta}
\end{figure}
We show the density ratio based on the numerical minimization of ${\cal E}(n,\beta)/n$, for Bogoliubov and Petrov's models,  
in Fig.~\ref{beta} for $g_{12} = -0.6g_{22}$ as a function of $g_{11}/g_{22}$. The $\beta$ values
are in very good agreement with the density ratio measured at the center of the finite-size self-trapped solutions obtained
by numerically solving the GP equations~(\ref{eeGPE}) and (\ref{oeGPE}) for Bogoliubov and Petrov's models, respectively, 
and start deviating from the simpler estimate $\sqrt{g_{22}/g_{11}}$ with an increase in asymmetry between intraspecies couplings.

For finite-size self-bound solutions within the SMA, the Lagrangian in Eq.~(\ref{Lag}) with $\Phi_\nu(x,t) = \alpha_\nu\Phi(x,t)$ and energy density in Eq.~(\ref{enden}) leads to a single Euler-Lagrange or GP equation
\begin{equation}
\label{GPE}
\begin{split}
    i\hbar\frac{\partial\Phi(x,t)}{\partial t} &= \biggl( -\frac{\hbar^2}{2m}\frac{\partial^2}{\partial x^2} + \frac{g_{11}\beta^2+g_{22}+2g_{12}\beta}{(\beta+1)^2} |\Phi(x,t)|^2 \\
    &\quad - \frac{3\sqrt{m(g_{11}\beta+g_{22})^3}}{4\hbar\pi\sqrt{(\beta +1)^3}}|\Phi|\mathcal{R}(p) \biggr) \Phi(x,t).
\end{split}
\end{equation}
To cast the above equation in dimensionless form, we first rescale the wavefunction as $\Phi = \Phi_0\tilde\Phi$ with
\begin{equation}
    \Phi_0 = \frac{3\sqrt{m(\beta +1)(g_{11}\beta+g_{22})^3}\mathcal{R}(p)}{4\pi\hbar(g_{11}\beta^2+g_{22}+2g_{12}\beta)},
    \label{phi0}
\end{equation}
and then consider
\begin{subequations}\label{units}
    \begin{align}
       l_0 &= \frac{4\hbar^2\pi\sqrt{(\beta + 1)(g_{11}\beta^2+g_{22}+2g_{12}\beta)}}{3m\sqrt{(g_{11}\beta+g_{22})^3}\mathcal{R}(p)},  \\
       t_0 &= \frac{16\hbar^3\pi^2(\beta +1)(g_{11}\beta^2+g_{22}+2g_{12}\beta)}{9m(g_{11}\beta+g_{22})^3[\mathcal{R}(p)]^2}, \\
       E_0 & = \frac{\hbar^2}{ml_0^2} = \frac{\hbar}{t_0} = \frac{9m(g_{11}\beta+g_{22})^3[\mathcal{R}(p)]^2}{16\hbar^2\pi^2(\beta + 1)(g_{11}\beta^2+g_{22}+2g_{12}\beta)},
    \end{align}
\end{subequations}
as units of length, time, and energy, respectively.
The dimensionless equation thus obtained is
\begin{equation}
    i\frac{\partial\tilde\Phi}{\partial \tilde t} =\Bigg( -\frac{1}{2}\frac{\partial^2}{\partial \tilde x^2} + |\tilde\Phi|^2 - |\tilde\Phi|\Bigg)\tilde\Phi,
    \label{dl_gpe}
\end{equation}
where $\tilde{x} = x/l_0$, $\tilde{t} = t/t_0$, and the number of particles $\tilde N = \int |\tilde \Phi|^2 d\tilde x$ is in units of
\begin{equation}
    N_0 = \Phi_0^2 l_0 = \frac{3\sqrt{(g_{11}\beta + g_{22})^3(\beta+1)^3}\mathcal{R}(p)}{4\pi\sqrt{(g_{11}\beta^2+g_{22}+2g_{12}\beta)^3}}.
    \label{N0}
\end{equation}
For equal intraspecies couplings, $g_{11}=g_{22} = g$, $\beta = 1$ and $p = g_{12}^2/g^2 - 1$, which fixes $l_0$, $t_0$, $E_0$,
and $\Phi_0$ in terms of coupling strengths.
The exact self-bound solution of Eq.~(\ref{dl_gpe}) is \cite{PhysRevLett.117.100401, PhysRevA.98.013631}
\begin{equation} \label{sb_soln}
\tilde \Phi(\tilde x,\tilde{t}) = -\frac{3\tilde{\mu} e^{-i\tilde{\mu} \tilde{t}}}{1+\sqrt{1+\frac{9\tilde{\mu}}{2}}\cosh(\sqrt{-2\tilde{\mu} \tilde{x}^2})},
\end{equation}
with the number of atoms 
\begin{equation}\label{Nmu}
\tilde N = \frac{4}{3}\left(\ln\frac{\sqrt{-9\tilde{\mu}/2}+1}{\sqrt{9\tilde{\mu}/2+1}}-\sqrt{-\frac{9\tilde{\mu}}{2}}\right).
\end{equation}
The form of Eq.~(\ref{dl_gpe}) matches the single GP equation derived by Astrakharchik and co-workers \cite{PhysRevLett.117.100401, PhysRevA.98.013631, PhysRevA.101.051601}, 
for which the mean-field energy $\cal E$ can be obtained from Eq.~(\ref{enden})
and respective units of various quantities from Eqs.~(\ref{phi0}), (\ref{units}), and (\ref{N0}) 
by substituting $\beta = \sqrt{g_{22}/g_{11}}$ and evaluating $\mathcal{R}(p)$ at $p=0$.  
The BdG equations corresponding to the linearization of the single GP equation (\ref{dl_gpe}) for this case have been used to calculate
density excitations in a pseudospinor mixture with equal intraspecies couplings \cite{PhysRevA.98.013631, PhysRevA.101.051601, PhysRevA.108.033312}. 
Equations~(\ref{sb_soln}) and (\ref{Nmu}) define the exact solution of the single-component reduced model, Eq.~(\ref{dl_gpe}), valid for all $N$ and permit two distinct regimes, separated by $\tilde N \sim 1$. 
For $\tilde N\ll 1$ (or $N\ll N_0$), the self-trapped solution has a bright solitonlike density profile, while for larger $\tilde{N}\gtrsim 10$ (or $N\gtrsim 10N_0$) a flat-top droplet emerges. 
As $\tilde{N}\rightarrow \infty$, one obtains an infinite-size flat-top droplet with peak density and the chemical potential getting saturated to $|\tilde{\Phi}|^2 =4/9$ and $\tilde{\mu}=-2/9$, respectively \cite{PhysRevA.98.013631}.
\subsection*{Excitation spectrum of a pseudospinor Bose-Bose droplet}
To describe the complete dynamics or excitation spectrum of a pseudospinor mixture as $\delta g$ increases from zero, the SMA needs to be
abandoned in favor of the dynamics described by the full model, viz. Eqs.~(\ref{eeGPE}) and (\ref{bdg}). In this direction, we cast these equations
in dimensionless form by expressing length, time, energy, and $\Phi_\nu$ in units of $\hbar^2/mg$, $ \hbar^3/mg^2$, $m g^2/\hbar^2 $, and $\sqrt{m g}/\hbar$, respectively, where $g=\sqrt{g_{11}g_{22}}$.

\begin{figure}[!htbp]
%\centering
\includegraphics[width=\columnwidth]{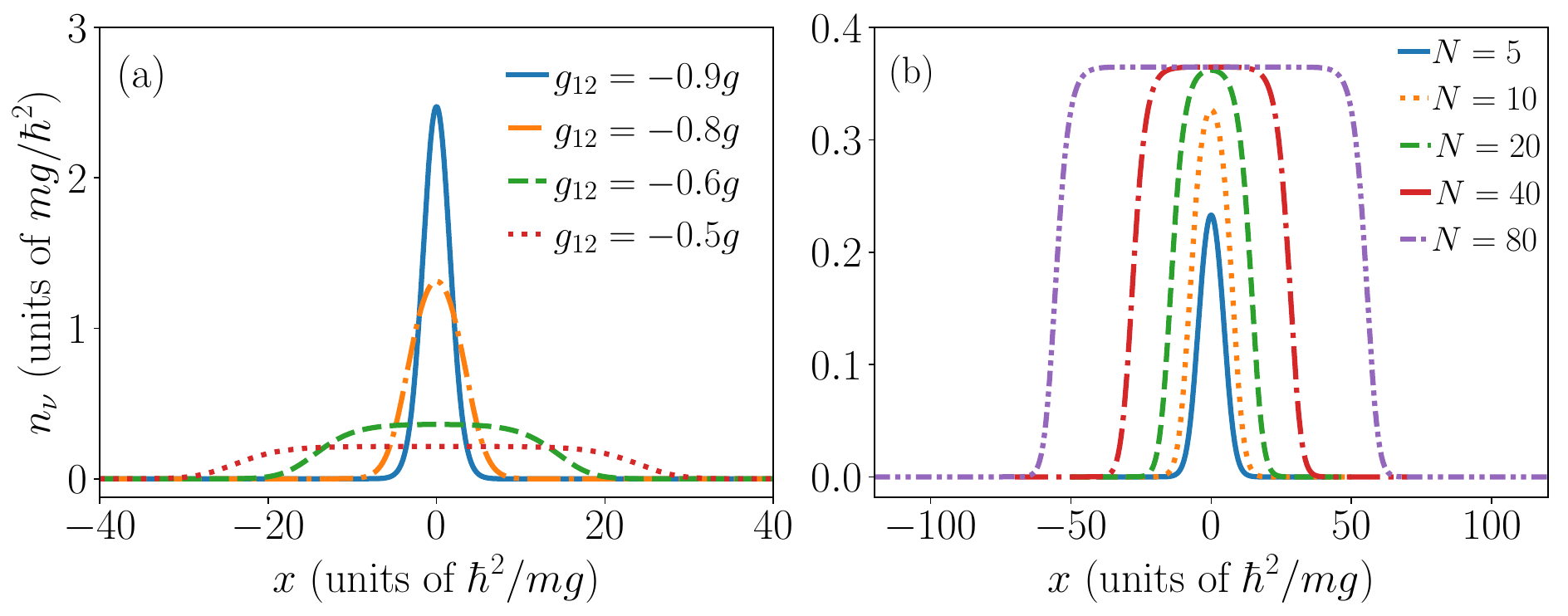}
\caption{Density profiles of the self-bound ground-state solutions of the Bogoliubov model, viz. Eq.~(\ref{eeGPE}), for equal
intraspecies coupling strengths, $g_{11} =g_{22} = g$. The component density is for (a) $N = 20$ and (b) $g_{12} = -0.6 g$, where $n_1(x) = n_2(x)$.}
\label{sp_br}
\end{figure}
\begin{figure*}[!htbp]
%\centering
\includegraphics[width= 0.9\textwidth]{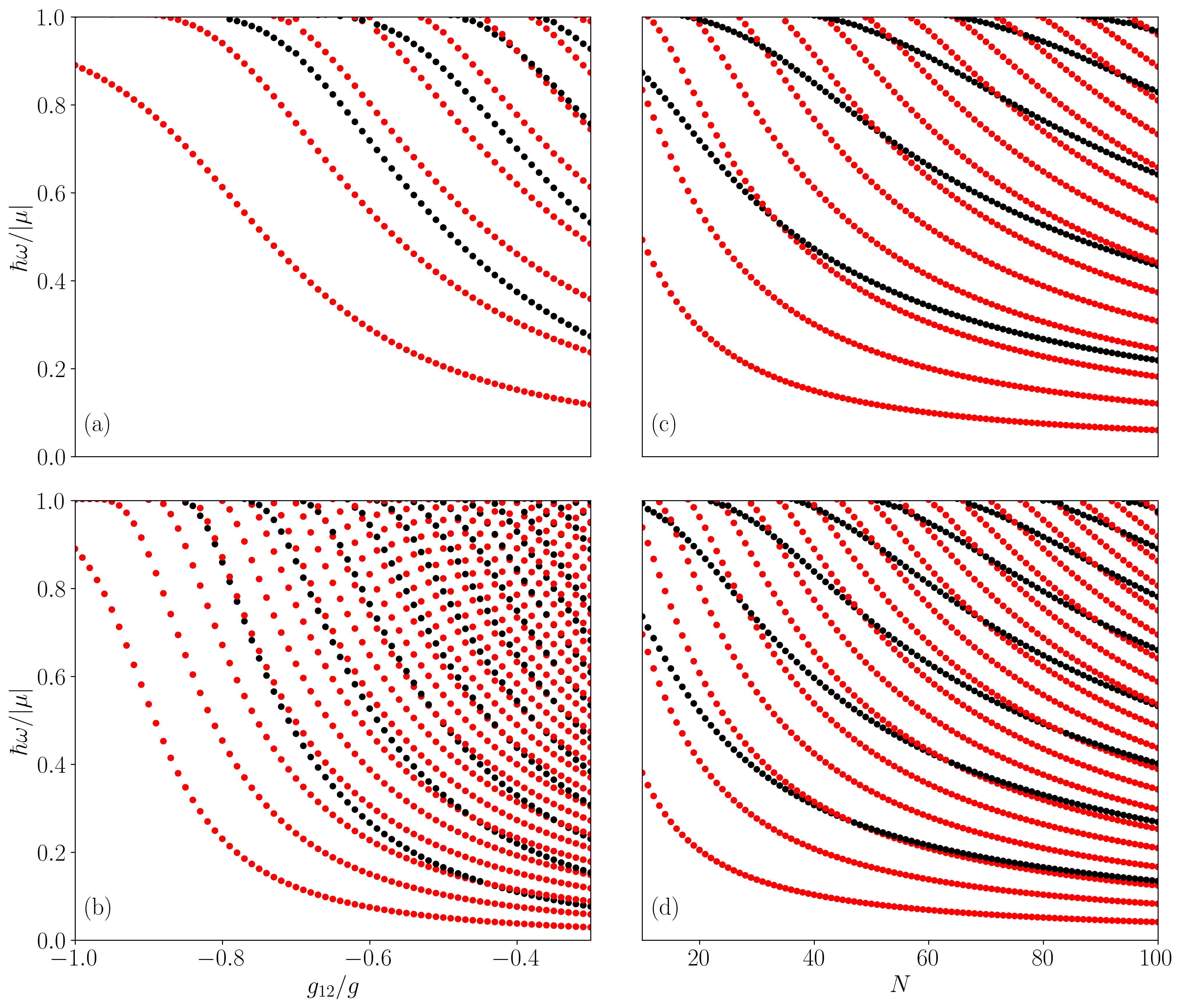}
\caption{Excitation energies calculated by solving the BdG equations~(\ref{bdg}) for the Bogoliubov model. Excitation energy is plotted as a function of $g_{12}/g$ for (a) $N = 20$ and (b) $N = 80$ and as a function of $N$ for (c) $g_{12} =-0.6g$ and (d) $g_{12} =-0.5g$.
Black (dark) and red (light) circles correspond to spin and density modes, respectively, identified from their $Q$ values. 
%\sout{Dashed blue and solid green lines denote the energies of density- and spin-breathing modes, respectively, obtained from the variational analysis [cf. Eqs. (\ref{omega_d}) 
%and (\ref{omega_s})]}.
The energies are in units of particle emission threshold $|\mu|$, and only the energies below the emission threshold are shown.}
\label{excitation_gN}
\end{figure*}
{\em Numerical results}. We consider $g_{11} = g_{22} = g>0$ for the pseudospinor mixture and 
solve Eq.~(\ref{eeGPE}) numerically using split-step imaginary-time propagation implemented via the Fourier pseudospectral method \cite{kaur2021fortress, *banger2022fortress, *banger2021semi} to obtain the ground state solutions. The method is implemented on a 1D spatial grid with spacing $\Delta x = 0.02$ and an extent $L = 360$.  
% where $L$ is chosen sufficiently large such that the density vanishes at the boundaries. 
The imaginary-time step considered is $0.1\times \Delta x^2$ and the convergence is declared when the change in $|\phi_\nu|$ between two consecutive iterations falls below a tolerance of $10^{-6}$.
% ; the box is enlarged until results are insensitive to $L$, which is essential for the extended flat-top and halo solutions.
Changes in the density profile of the ground state of the mixture with an increase in $g_{12}$ for fixed $N$ or with an increase in $N$ 
for fixed $g_{12}$ are shown in Figs.~\ref{sp_br}(a) and \ref{sp_br}(b), respectively, and are in agreement with the observations made in Refs.~\cite{PhysRevLett.117.100401, 10.21468/SciPostPhys.19.5.133}. 
Using the ground state solutions, we calculate the full excitation spectrum of the mixture by solving the BdG equations~(\ref{bdg}) numerically for the Bogoliubov model
using a basis expansion method \cite{roy-2020, PhysRevA.111.023311}. Specifically, the energy eigenstates of a particle confined to a box of length $L$ 
serve as a suitable basis truncated to the lowest $500$ eigenstates; low-lying eigenvalues are converged with respect to the basis size. Self-binding is confirmed by at least one collective mode below the particle emission threshold $-\mu$. For the solutions presented in this work, all physical eigenvalues are real, 
implying no dynamical instability. The BdG operator yields $\pm \omega_l$ eigenvalue pairs of opposite norm $\sum_{\nu}\int dx\left[|u_\nu^l(x)|^2 -|v_\nu^{l}(x)|^2\right]$; the positive eigenvalues have positive norm and the negative eigenvalues have negative norm, which
is indicative of the energetic stability of the droplet solutions \cite{pethick_smith_2008}. These are consistent with the fact that droplets reside in the mean-field stable regime. The only zero-norm modes in the spectrum are three zero-energy modes: Two of these correspond to the Goldstone modes due to the breaking of U(1) gauge symmetry, while one arises from self-trapping \cite{Zin_2021}. For the sake of brevity, we have not included the zero-energy modes in the spectrum. These modes also serve as a self-consistency check to verify the accuracy of our numerical calculations.  
We characterize the density or spin character of the $l$th excitation mode by
a measure $Q_l$ defined as $Q_l = S^l_-/S^l_+$, where  $S_{\pm}^l = \int [\{\delta n^l(x)\}^2 \pm \{\delta s^l(x)\}^2]dx$ \cite{l529-9544, PhysRevA.108.043310}.
Here $\delta n^l(x) =\delta n^l_1(x)+\delta n^l_2(x)$ and $\delta  s^l(x) =  \delta n^l_1(x)-\delta n^l_2(x)$,
where $\delta n_\nu^l(x) = 2\mathrm{Re}\left[\phi_\nu \left(u_\nu^{l*} - v_\nu^l\right)\right] $ is the density fluctuation in the component $\nu$ generated by the $l$th Bogoliubov mode.

The excitation spectrum as a function of $g_{12}/g$ for $N = 20$ and $N = 80$ are shown in Figs.~\ref{excitation_gN}(a) and \ref{excitation_gN}(b), respectively.
With an increase in $g_{12}$, i.e., an increase in net mean-field repulsion, all the excitation energies monotonically decrease. For $-g< g_{12}\lesssim-0.775 g$ for $N = 20$
and $-g<g_{12}\lesssim -0.85 g$ for $N = 80$, all the excitations below the particle emission threshold are density excitations with $Q = 1$, the lowest of which is the density
breathing mode. For $g_{12}\gtrsim -0.775g~(-0.85g)$ for $N=20~(80)$, the spin channel is sufficiently softened, leading to the appearance of spin excitations below the particle emission threshold with $Q= -1$.
Similar to the density channel, the lowest-lying spin excitation is the (spin) breathing mode.
We also examine the excitation spectrum as a function of $N$ for fixed coupling strengths [see Fig.~\ref{excitation_gN}(c) for $g_{12} = -0.6g$ and Fig.\ref{excitation_gN}(d) for $g_{12} = -0.5 g$]. 
Here too there is a monotonic decrease in excitation energies with more spin excitations appearing below the 
particle emission threshold as $N$ increases. For $g_{11}= g_{22}$, the density and spin channels are fully decoupled, as evidenced by the modes 
with $Q = 1$ or $-1$, where the density and spin modes correspond to in-phase and out-of-phase modes, respectively.
\begin{figure}[!htbp]
\centering
\includegraphics[width=\columnwidth]{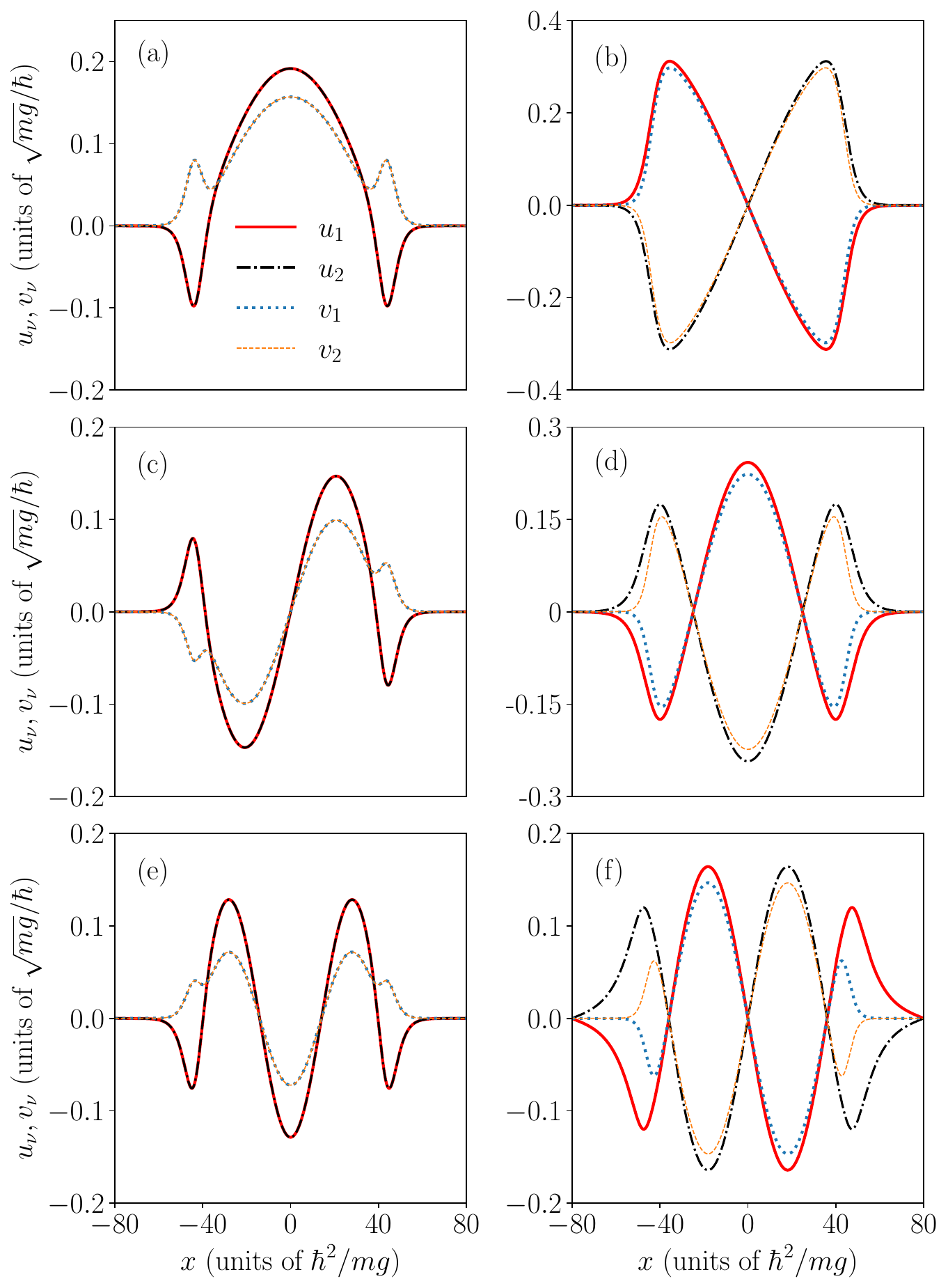}
\caption{Quasiparticle amplitudes of the three lowest-lying density and spin modes
for $N = 60$ and $g_{12} = -0.6 g$: (a) first, (c) second, and (e) third density mode and (b) first, (d) second, and (f) third spin mode. Density modes correspond to in-phase oscillations with $u_1 = u_2$ and $v_1=v_2$, whereas spin modes correspond to out-of-phase oscillations with $u_1 = -u_2$ and $v_1 = -v_2$.}
\label{uv}
\end{figure}
This is demonstrated by the quasiparticle amplitudes $(u_\nu,v_\nu)$ of the three lowest-lying density modes in the left column of Fig.~\ref{uv} and spin modes in the right column of Fig.~\ref{uv}
for $g_{12} = -0.6$ and $N = 60$; density modes have $u_1 = u_2$ and $v_1 = v_2$, whereas for spin modes $u_1 = -u_2$ and $v_1= -v_2$. 
\begin{figure}[!htbp]
\centering
\includegraphics[width=\columnwidth]{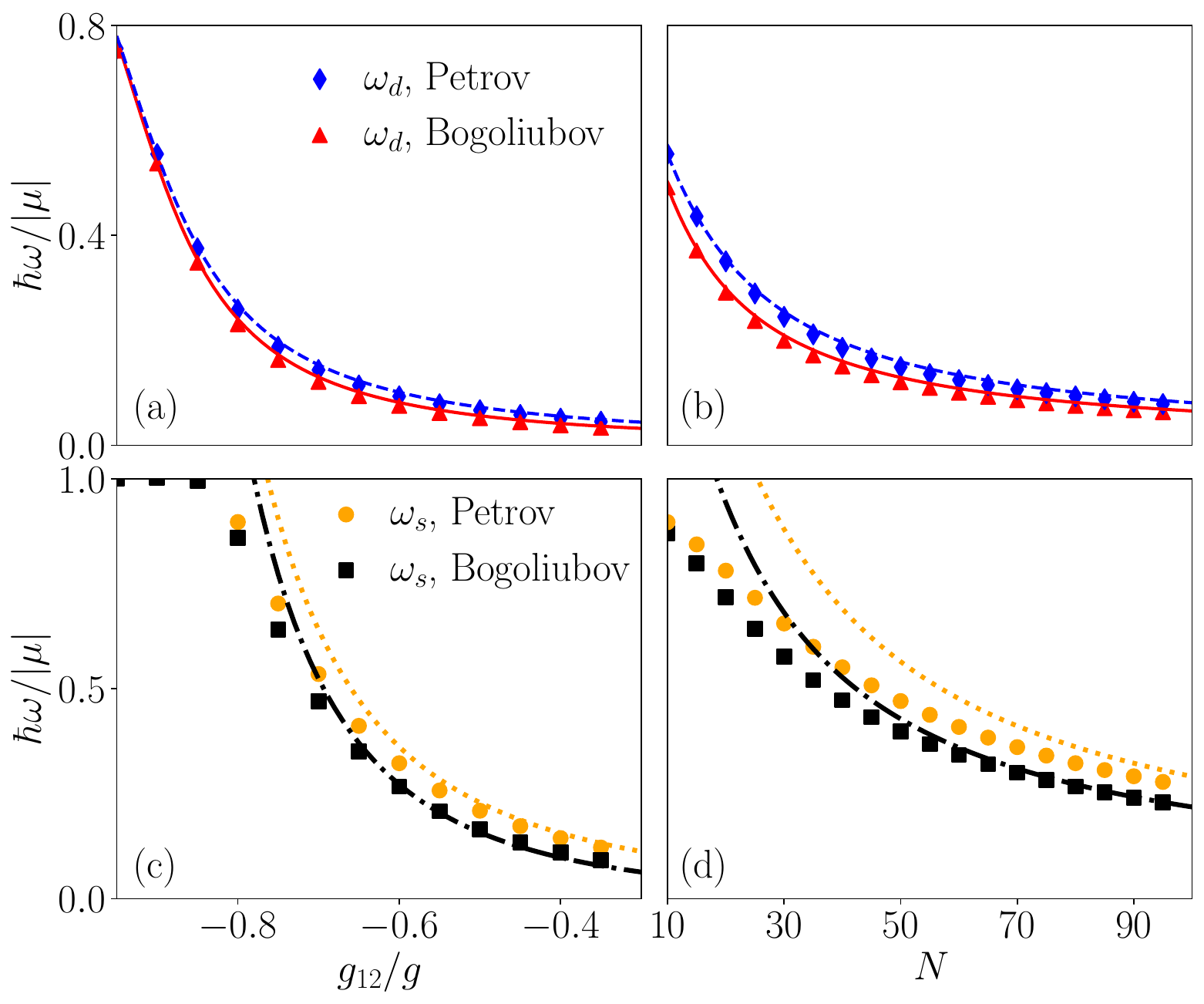}
\caption{Comparison of the (a) and (b) density-breathing and (c) and (d) spin-breathing mode energies predicted by the Petrov and Bogoliubov models. Excitation energy is plotted (a) and (c) as a function of $g_{12}/g$ for $N=80$ and (b) and (d) as a function of $N$ for $g_{12}/g=-0.6$. Numerical BdG results (symbols) are compared with variational predictions (lines of the corresponding colors), demonstrating that the variational ansatz accurately captures the essential physics of the collective modes. The variational results are obtained from Eqs.~(\ref{omega_d}) and (\ref{omega_s}): The red solid (blue dashed) curve represents the density-breathing mode in the Bogoliubov (Petrov) model, while the black dash-dotted (orange dotted) curve represents the spin-breathing mode in the Bogoliubov (Petrov) model.
}
\label{pet_bog}
\end{figure}
For $g_{11} = g_{22}$, $\mathcal {R}(p)$ monotonically decreases from $4/3$ to $2\sqrt{2}/3$ as $g_{12}$ increases from $-g$ to $0$ \cite{10.21468/SciPostPhys.19.5.133}.
This effectively reduces the strength of the attractive LHY correction, resulting in flatter droplets within Bogoliubov theory vis-à-vis Petrov's.
We find that excitation energies within Bogoliubov theory are perceptibly lower than those obtained in Petrov's modeling. This is illustrated by the density and spin breathing modes calculated for the two models, which are plotted as a function of \( g_{12} \) in Figs. \ref{pet_bog}(a) and \ref{pet_bog}(c), respectively. Additionally, the same modes are shown as a function of \( N \) in Figs. \ref{pet_bog}(b) and \ref{pet_bog}(d).

{\em Variational analysis}.
To further understand the density- and spin-breathing modes, we consider the variational ansatz given by
\begin{equation} \phi_{\nu}(x,t) = \sqrt{\frac{N}{2\sqrt{\pi}\sigma_\nu(t)}} \exp \left[ -\frac{x^2}{2\sigma_\nu(t)^2} + i \beta_\nu(t)x^2 \right], 
\label{ansatz} 
\end{equation} 
where $\sigma_\nu(t)$ and $\beta_\nu(t)$, which denote the width of the self-bound solution and the chirp, respectively, are the time-dependent 
variational parameters. 
The Lagrangian of the system is 
\begin{subequations}\label{Lagrangian}
\begin{align} L =& \int dx \sum_{\nu}\frac{i}{2}\left(\phi_\nu^* \frac{\partial \phi_\nu}{\partial t}- \phi_\nu\frac{\partial \phi_\nu^*}{\partial t}\right) -E \label{lagrangian},\\
                     =&-\frac{N}{4} \sum_{\nu=1,2} \sigma_\nu^2 \dot{\beta}_\nu-E
                   % =& L_{\rm {K}} - E_{\text{MF}} - E_{\text{LHY}} ,                    
\end{align} 
\end{subequations}
where $E = E_{\rm kin}+E_{\rm MF} + E_{\rm LHY}$ is the total energy of the system with $E_{\rm kin}$, $E_{\rm MF}$ and $E_{\rm LHY}$ denoting the kinetic, mean-field, and beyond-mean-field interaction contributions, respectively. These are given as 
\begin{align}
    E_{\rm kin} &= \int dx \left(\frac{1}{2}\sum_{\nu=1,2}\left|\frac{d\phi_\nu}{dx}\right|^2 \right),\\ 
    & = \frac{N}{4} \sum_{\nu=1,2} \left( 2\sigma_\nu^2 \beta_\nu^2 + \frac{1}{2\sigma_\nu^2} \right),\nonumber\\
    E_{\rm MF}&= \int \frac{1}{2}\sum_{\nu \nu'}g_{\nu \nu'}n_\nu n_{\nu'}dx,\nonumber \\
    &=\frac{N^2}{16\sqrt{\pi}} \left( \frac{\sqrt{2} g(\sigma_1+\sigma_2)}{\sigma_1 \sigma_2} + \frac{4 g_{12}}{\sqrt{\sigma_1^2+\sigma_2^2}} \right) ,\label{E_MF}\\
    E_{\rm LHY} &= -\int dx \frac{g^{3/2}}{2\pi} \left(\sum_{\nu=1,2} |\phi_\nu|^2\right)^{3/2}{\cal R}(p)\label{L_LHY_int}.
\end{align}
 %\sout {To make it analytically tractable} 
To evaluate the integration in $E_{\rm LHY}$, we express widths as $\sigma_{1}(t)=\bar\sigma(t)+\epsilon(t)$ and $\sigma_{2}(t)= \bar\sigma(t)-\epsilon(t)$, where $\bar\sigma(t) =[\sigma_1(t)+\sigma_2(t)]/2 $ and $\epsilon(t) = [\sigma_1(t)-\sigma_2(t)]/2$. As $|\epsilon(t)|\ll\bar\sigma(t)$, we write the integrand as a single-variable Taylor series expansion 
in $\epsilon(t)$ about $\epsilon(t) =0$ and retain the terms up to second order in $\epsilon(t)$ (see Appendix B).
The $E_{\rm LHY}$ thus obtained is 
\begin{align}
E_{\rm LHY} \approx& \frac{2 (g N)^{3/2}}{3\sqrt{6}\pi ^{5/4} (\sigma_1+\sigma_2)^{5/2}}\Bigg[\mathcal{C}_1 (\sigma_1+\sigma_2)^2\nonumber\\
            &+\frac{3}{4} \mathcal{C}_2  (\sigma_1-\sigma_2)^2\Bigg],
\end{align}
 where $\mathcal{C}_1$, and $\mathcal{C}_2$ are defined in Eq.~(\ref{c1c2}) in Appendix B.
Using the Lagrangian, the Euler-Lagrange equation for the variational parameter $\beta_{\nu}(t)$,
\begin{equation}
    \frac{\partial L}{\partial \beta_\nu} = \frac{d}{dt}\left(\frac{\partial L}{\partial \dot \beta_\nu}\right)
\end{equation}
yields $\beta_\nu(t)=\dot{\sigma}_\nu(t)/2\sigma_\nu(t)$. Similarly, the Euler-Lagrange equation of $\sigma_{\nu}(t)$
is obtained. For more details, we refer the reader to Eq.~(\ref{condensate_width}) and the follow-up discussion in Appendix B. 
We linearize the two coupled Euler-Lagrange equations for variables $\sigma_\nu(t)$
about the equilibrium widths $\sigma_1^{\rm eq} = \sigma_2^{\rm eq} =\sigma$ %{\color{BrickRed} \sout{(see Appendix)}}
to arrive at the coupled equations for fluctuations in widths about the equilibrium value
\begin{subequations}
\begin{align}
    \delta \ddot \sigma_1(t) + \omega_{11}^2\delta \sigma_1(t)  + \omega_{12}^2\delta \sigma_2(t)  = 0, \\
    \delta \ddot \sigma_2(t) + \omega_{21}^2\delta \sigma_1(t)  + \omega_{22}^2\delta \sigma_2(t)  = 0,
\end{align} \label{linearize}
\end{subequations}%where {\color{red}$\omega_{\nu\nu'}^2 = - \frac{\partial \ddot \sigma_\nu(t)}{\partial \sigma_{\nu'}(t)}$} at equilibrium ($\sigma_1(0)=\sigma_2(0)=\sigma$). %Since, we have 
where $ \omega_{11}^2 = \omega_{22}^2 = \omega_{1}^2$ and $ \omega_{12}^2 = \omega_{21}^2 = \omega_{2}^2$.
The equilibrium width $\sigma$ satisfies 
\begin{equation}
\frac{1}{\sigma ^3}+\frac{N \left(g_{12} +g\right)}{2 \sqrt{2 \pi } \sigma ^2}+\frac{ g ^{3/2}\sqrt{N}\mathcal{C}_1}{3 \sqrt{3} \pi ^{5/4} \sigma ^{3/2}}=0,
\end{equation}
and $\omega_1^2$ and $\omega_2^2$ have the forms
\begin{subequations}
\begin{align}
\omega_{1}^2 &= \frac{3}{\sigma^4} + \frac{(4g+g_{12})N}{4\sqrt{2 \pi} \sigma^3} + \frac{ g^{3/2}\sqrt{N}}{4\sqrt{3} \pi^{5/4} \sigma^{5/2}} \Big( \mathcal{C}_1 + 2\mathcal{C}_2 \Big), \label{omega11}\\
\omega_{2}^2 &= \frac{3 g_{12} N}{4\sqrt{2\pi}\sigma^3} + \frac{g^{3/2} \sqrt{N} }{4\sqrt{3} \pi^{5/4} \sigma^{5/2}} \Big( \mathcal{C}_1 - 2\mathcal{C}_2 \Big).
\label{omega22}
\end{align}
\end{subequations}
Using $\delta \sigma_1(t) = \delta \sigma_2(t)$ for the density-breathing mode and $\delta \sigma_1(t)= - \delta \sigma_2(t)$ for spin-breathing
mode in Eqs.~(\ref{linearize}a) and (\ref{linearize}b), the frequencies of the density-breathing mode ($\omega_d$) and spin-breathing mode ($\omega_s$) are 
%\sout{$\omega_d^2 = (\omega_1^2+\omega_2^2)$ and $\omega_s^2 = (\omega_1^2-\omega_2^2)$.}
\begin{subequations}
\begin{align}
\omega_d^2 &= \omega_1^2+\omega_2^2
=\frac{3}{\sigma^4}
+\frac{N(g+g_{12})}{\sqrt{2\pi}\sigma^3}
+\frac{\sqrt{N}g^{3/2}\mathcal{C}_1}
{2\sqrt{3}\,\pi^{5/4}\,\sigma^{5/2}},\label{omega_d}\\
\omega_s^2 &= \omega_1^2-\omega_2^2
=\frac{3}{\sigma^4}
+\frac{N(2g-g_{12})}
{2\sqrt{2\pi}\sigma^3}
+\frac{\sqrt{N}\,g^{3/2}\mathcal{C}_2}
{\sqrt{3}\pi^{5/4}\sigma^{5/2}}\label{omega_s}.
\end{align}
\end{subequations}
 
The variational estimates of 
density- and spin-breathing modes are in very good agreement with the numerical results, especially for larger
numbers of bosons [see Figs.~\ref{pet_bog}(b) and \ref{pet_bog}(d)].  We emphasize that the variational treatment targets only the lowest excitation, breathing mode, 
of each sector (spin and density), while the BdG results (see Fig.~\ref{excitation_gN}) show the complete excitation spectrum below the particle-emission threshold. Beyond reproducing the numerical spectrum, the variational treatment makes transparent the physical origin of the density- and spin-breathing modes as in-phase and out-of-phase width oscillations, providing an analytical counterpart to the BdG results.

\section{Scalar Bose-Bose Mixture}
\label{Sec-III}
\begin{figure*}[!htbp]
\centering
\includegraphics[width=0.9\textwidth]{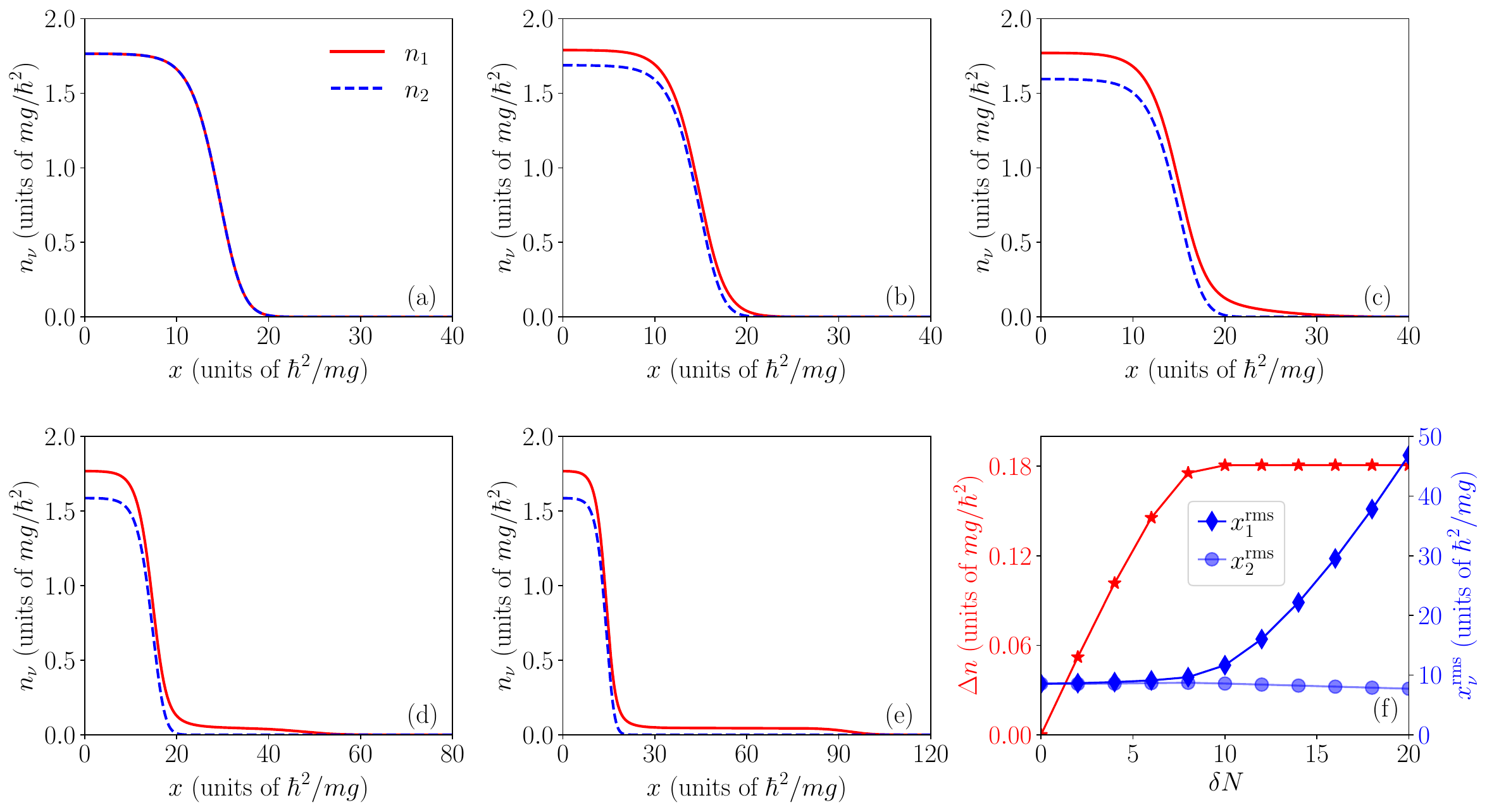}
\caption{Density profiles of scalar mixtures with $g_{12} = -0.8g$ and $N = 100$ for (a) $\delta N=0$, (b) $\delta N=4$, (c) $\delta N=8$, (d) $\delta N=10$, and (e) $\delta N=14$. 
(f) Difference in the densities of the two components at $x = 0$, $\Delta n = n_1(x=0)-n_2(x=0)$, and the rms sizes of the components, $x_\nu^{\rm rms}$, as a function of $\delta N$.
As the densities are symmetric about $x=0$, i.e. $n_\nu(x) = n_{\nu}(-x)$, densities are shown only for $+x$ values. }
\label{den_2com}
\end{figure*}
We now study a scalar Bose-Bose mixture in which the number of atoms in each species, $N_\nu$, is independently fixed.
To examine the excitation spectrum as a function of particle imbalance  $\delta N = N_1 - N_2$,
we consider the mixture with equal intraspecies coupling strengths ($g_{11}=g_{22}=g$) and $g_{12} = -0.8g$.
In Fig.~\ref{den_2com} we show the densities of two components for different $\delta N$ with $N = 100$.
The densities are symmetrical about $x=0$ for all $\delta N$. For $N_1=N_2$, the densities of the two components overlap as shown in Fig.~\ref{den_2com}(a). As $\delta N$ is increased, the difference in the peak densities at $x=0$ increases for small particle imbalances [see Figs.~\ref{den_2com}(a)-\ref{den_2com}(c)]. The additional atoms are accommodated by the droplet
without any significant increase in the root mean square (rms) sizes of the two components for $\delta N< 8$. On a further increase in $\delta N$, 
the difference in the peak densities and the rms size of the minority component do not change much, whereas the rms size of the majority component
grows exponentially due to the excess (unbound) atoms sandwiching the droplet.
The densities for $\delta N =10$ and  $\delta N =14$ are shown in Figs.~\ref{den_2com}(d) and \ref{den_2com}(e), respectively.
The variation in the peak density difference, $n_1(x= 0) -n_2(x=0)$, and the rms sizes of the components $x_\nu^{\rm rms}$
as a function of $\delta N$ are shown in Fig.~\ref{den_2com}(f). Similar changes in the ground-state features have been
observed in recently studied particle-imbalanced 3D quantum droplets \cite{PhysRevResearch.5.033167, PhysRevResearch.6.013209}. 

\begin{figure}[!htbp]
\centering
\includegraphics[width=\columnwidth]{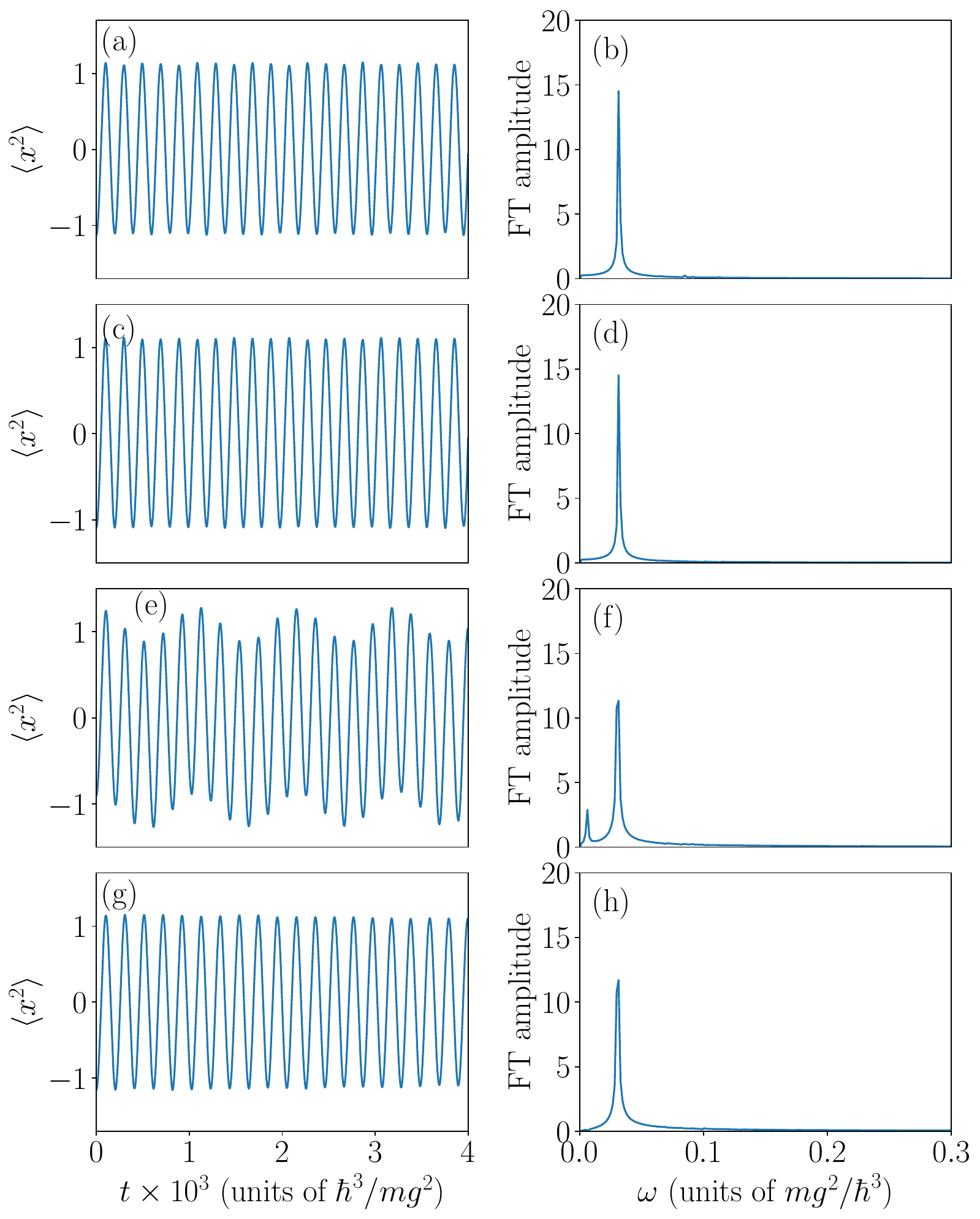}
\caption{Temporal oscillations in the mean-square sizes $\langle x^2 \rangle_\nu= \int x^2 |\Phi_\nu(x,t)|^2 dx/N_\nu$ of (a) major ($\nu =1$) and (c) minor ($\nu =2$) components for $N = 100$, $g_{12} = -0.8g$ and a particle imbalance $\delta N = 4$. The oscillations are centered around zero 
by subtracting their equilibrium mean values. The corresponding Fourier transform (FT) amplitudes are shown in (b) and (d). The oscillations in $\langle x^2\rangle_\nu$ are shown for (e) major and (g) minor components, with particle imbalance $\delta N = 10$ alongside
their respective Fourier transforms in (f) and (h). Here $\langle x^2 \rangle_\nu$ is measured in units of  $\hbar^4/m^2g^2$.}
\label{dynamics_scalar}
\end{figure}
\begin{figure}[!htbp]
\centering
\includegraphics[width=\columnwidth]{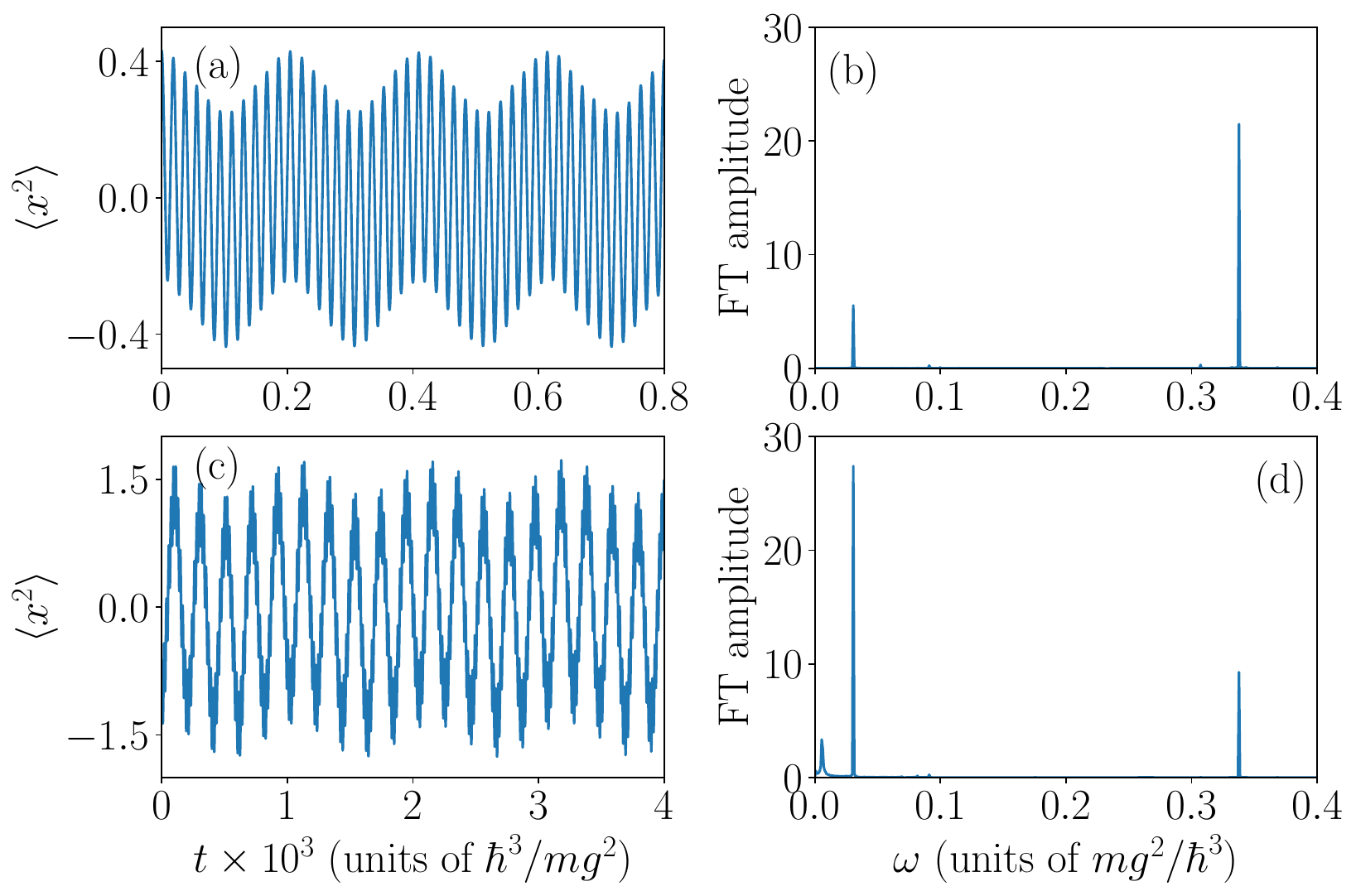}
\caption{(a) Mean-subtracted temporal oscillations of the first term on the right-hand side of Eq.~(\ref{x1_ms}), for $N=100$, $\delta N = 10$, and $g_{12} = -0.8g$ 
with (b) the amplitude of its Fourier transform. (c) Identical oscillations of the other two terms in Eq.~(\ref{x1_ms}), with (d) their Fourier transforms.}
\label{dynamics_major}
\end{figure} 
We first study the breathing oscillations of imbalanced droplets by tracking the real-time evolution of ground states generated by a perturbed Hamiltonian. 
More precisely, we perturb the intraspecies coupling $g$ by changing it to   $1.001g$ at $t=0$ 
and examine the oscillations in mean-square sizes, $\langle x^2\rangle_\nu(t) = \int x^2|\Phi_\nu(x,t)|^2 dx/N_\nu$, of the two components. For $\delta N < 8$, $\langle x^2\rangle_\nu(t)$ oscillate with a single 
frequency, as shown in Figs.~\ref{dynamics_scalar}(a)-\ref{dynamics_scalar}(d) for $\delta N = 4$. This represents the density-breathing mode of the droplet.  
For $\delta N \geq 8$, $\langle x^2\rangle_1(t)$ oscillates with a distinctive beat pattern in contrast 
to $\langle x^2\rangle_2(t)$, as seen in Figs.~\ref{dynamics_scalar}(e) and \ref{dynamics_scalar}(g) for $\delta N = 10$. The Fourier transform of $\langle x^2\rangle_\nu(t)$ [see Figs.~\ref{dynamics_scalar}(f) and \ref{dynamics_scalar}(h)]
confirms two frequencies in $\langle x^2\rangle_1(t) $, the larger of which matches the oscillation frequency of the minority component. The ground-state solution for $\delta N \geq 8$ has two distinct regions: an overlap or core region corresponding to the self-bound droplet
and an outer region corresponding to the halo of the majority component's atoms. This suggests that the beat pattern in the oscillations
of $\langle x^2\rangle_1(t)$ may correspond to two different length scales, corresponding to the size of the core or droplet proper and the majority component's
overall size, which includes the outer unbound gas in which the droplet now lies immersed. To shed more light on this, we
approximate the overlap extent of two components by the rms size of the minority component $x_2^{\rm rms} = \sqrt{\langle x^2\rangle_2(t=0)}$ and 
split  $\langle x^2\rangle_1(t)$ as
\begin{align}
 \langle x^2\rangle_1(t) =&  \frac{1}{N_1}\left[\int_{-x_2^{\rm rms}}^{x_2^{\rm rms}} x^2|\Phi_1(x,t)|^2 dx \right.\nonumber\\
                 & \left.+\int_{-\infty }^{-x_2^{\rm rms}} x^2|\Phi_1(x,t)|^2 dx + \int_{x_2^{\rm rms} }^{\infty} x^2|\Phi_1(x,t)|^2 dx\right]
                 \label{x1_ms}.
\end{align}
The first term on the right-hand side of Eq.~({\ref{x1_ms}}) corresponds to the overlap region 
and is plotted in Fig.~\ref{dynamics_major}(a) and its Fourier transform in Fig.~\ref{dynamics_major}(b), showing that it oscillates with two frequencies, with the smaller of the two matching the oscillation frequency of the minority component and the other corresponding to the chemical potential
of the minority component. 
We identify the smaller frequency to be associated with the density-breathing mode of the droplet for $\delta N\geq8$.
The oscillations in the other two terms plotted in Fig.~\ref{dynamics_major}(c) and its Fourier transform in Fig.~\ref{dynamics_major}(d) 
contain three frequencies, two of which match the oscillations in $\langle x^2\rangle_1(t)$ in Fig.~\ref{dynamics_scalar}(e) and a third, the largest of the three frequencies, 
corresponding to the chemical potential of the minority component.

\begin{figure*}[!htbp]
\centering
\includegraphics[width=0.9\textwidth]{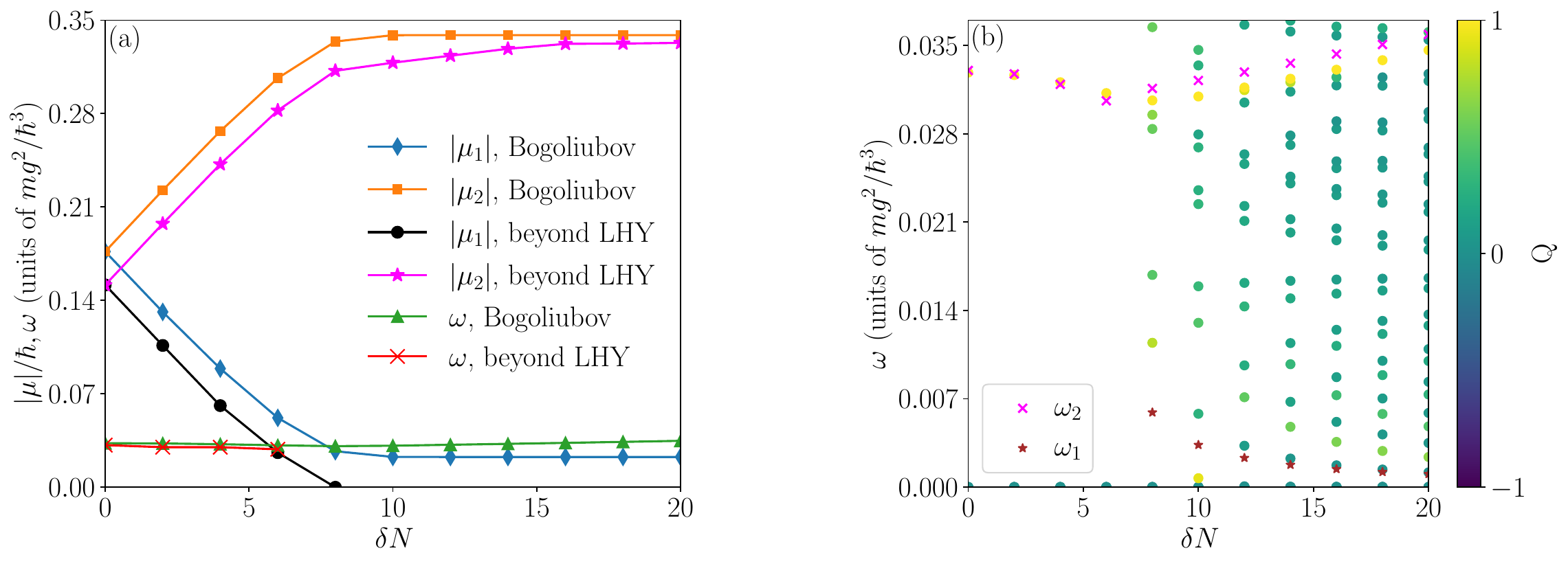}
\caption{
(a) Plot of $|\mu_1|/\hbar$, $|\mu_2|/\hbar$, and the frequency of the density-breathing mode ($\omega$) as a function of $\delta N = N_1 - N_2$
of a scalar mixture with $N=100$ and $g_{12}=-0.8g$. For $\delta N \ge 8$, the Bogoliubov theory leads to the density-breathing mode's energy exceeding $|\mu_1|$. 
The beyond-LHY theory leads to reduced $|\mu_\nu|$ with $|\mu_1| \approx 0$ for $\delta N\ge8$. Using the beyond-LHY theory, the breathing mode is calculated
only for $\delta N<8$ as the solutions are not bounded by any finite solution domain $L$ for $\delta N\ge 8$ as is discussed in Sec. \ref{Sec-IV}. 
(b) Low-lying excitation energies of the scalar mixture with $N=100$ and $g_{12}=-0.8g$ along with variational estimates $\omega_1$ and $\omega_2$ [density-breathing mode $\omega$ in (a)] as given in Eqs. (\ref{omega1}) and (\ref{omega2}).} 

\label{exc_2com}
\end{figure*}
In Fig.~\ref{exc_2com}(a) we plot the negative of the chemical potentials of two components and the density breathing mode frequency as a function of particle imbalance. With an increase in $\delta N$, the absolute value of the chemical potential of the minority (majority) component increases (decreases) until $\delta N<8$, beyond which the chemical potentials $\mu_\nu$ do not change.
The density-breathing mode is below the particle-emission thresholds for both components for $\delta N<8$, whereas for $\delta N \geq 8$, the density-breathing mode's frequency becomes larger than the particle-emission threshold for the majority component. Similar observations have been made for particle-imbalanced 3D droplets \cite{PhysRevResearch.5.033167, PhysRevResearch.6.013209}. We further calculated the complete
excitation spectrum of the mixture as a function of $\delta N$ by solving the BdG equations~(\ref{bdg}). From the spectrum shown in Fig.~\ref{exc_2com}(b), we find that for $\delta N<8$, the density-breathing mode with $Q=1$ decreases with increasing population imbalance and is the lowest-lying nonzero-energy excitation with the same frequency as identified from the real-time dynamics. The situation is drastically different for $\delta N\geq 8$ when there are many nonzero excitations all below the particle-emission threshold of the majority component, i.e., with $\omega<|\mu_1|$. The density mode gradually hardens as $\delta N$ is increased, with its frequency matching with the results from the real-time dynamics. The lowest nonzero frequency matches with the smaller of the two frequencies in the oscillations of $\langle x^2\rangle_1(t)$ for $\delta N\geq 8$ [see Figs.~\ref{dynamics_scalar}(e) and \ref{dynamics_scalar}(f)].  
The spectrum shown in Fig.~\ref{exc_2com}(b) therefore validates the real-time dynamics and the identification of the density-breathing mode in the preceding paragraph. 

{\em Variational analysis}. To further vindicate the role of two length scales on the breathing dynamics of the system,  we construct a variational analysis. We consider the variational ansatz given by 
\begin{subequations}
    \begin{align}
         \phi_{1}(x,t) &= 
         \begin{cases}
         \phi_1^{\rm out}(x,t), |x|\gtrapprox 3\sigma_2(t)\nonumber\\
         \phi_1^{\rm in}(x,t), |x|\lessapprox 3\sigma_2(t)\nonumber\\
         \end{cases}\\
         &=\begin{cases}
         A_1 \sqrt{\frac{N_1}{\sqrt{\pi}\sigma_1(t)}} e^{ \left[ -\frac{x^2}{2\sigma_1(t)^2} + i \beta_1(t)x^2 \right]},|x|\gtrapprox 3\sigma_2(t)\\
         A_2 \sqrt{\frac{N_1}{\sqrt{\pi}\sigma_2(t)}} e^{\left[ -\frac{x^2}{2\sigma_2(t)^2} + i \beta_2(t)x^2 \right]}, |x|\lessapprox 3\sigma_2(t)\\
         \end{cases}\\
         \phi_{2}(x,t) &= \sqrt{\frac{N_2}{\sqrt{\pi}\sigma_2(t)}} \exp \left[ -\frac{x^2}{2\sigma_2(t)^2} + i \beta_2(t)x^2 \right],
    \end{align}
\end{subequations}
where $\sigma_2(t)$ is a measure of the size of the overlap region or of the self-bound droplet, and $\sigma_1(t)$ is the overall size of the majority component, including the outer ``shell" of the unbound gas consisting solely of atoms of the majority component.
We assume that $\sigma_1(t)\gg \sigma_2(t)$ to make the integrals in the Lagrangian~(\ref{lagrangian}) tractable 
%{\color {BrickRed}\sout {for example, $L_{\rm LHY}$ in Eq.~(\ref{L_LHY}) we can be approximated as} 
and approximate the Lagrangian as%}  
\begin{align}
L=& \int_{-\infty}^{\infty} \mathcal {L}(\phi_1, \phi_2) dx\nonumber, \\
        \approx& \int_{-\infty}^{\infty} \mathcal {L}(\phi_1^{\rm in}, \phi_2) dx + \int_{-\infty}^{\infty} \mathcal {L}(\phi_1^{\rm out}, \phi_2 \approx 0 ) dx.
\end{align}
{ Then the Lagrangian for the scalar mixture is
\begin{align}
    L &= -\frac{N_w \sigma _1^2 \dot \beta _1+\tilde{N} \sigma _2^2 \dot \beta _2}{2} - E_{\rm kin} - E_{\rm MF} - E_{\rm LHY},
\end{align}
where
\begin{subequations}\label{E_comp_sm}
    \begin{align}
         E_{\rm kin} &= \frac{N_w \left(4 \beta _1^2 \sigma _1^4+1\right)}{4 \sigma _1^2}+\frac{\tilde{N} \left(4 \beta _2^2 \sigma _2^4+1\right)}{4 \sigma _2^2},\\
        % E_{\rm MF} &= \frac{N_w \left(4 \beta _1^2 \sigma _1^4+1\right)}{4 \sigma _1^2}+\frac{\tilde{N} \left(4 \beta _2^2 \sigma _2^4+1\right)}{4 \sigma _2^2}\nonumber\\&+\frac{g N_w^2}{2 \sqrt{2 \pi } \sigma _1}+\frac{g\tilde{N}^2}{2 \sqrt{2 \pi } \sigma _2}+\frac{A_2^2N_1N_2(g_{12}-g)}{\sqrt{2 \pi } \sigma _2},\\
          E_{\rm MF} &= \frac{g N_w^2}{2 \sqrt{2 \pi } \sigma _1}+\frac{g\tilde{N}^2}{2 \sqrt{2 \pi } \sigma _2}+\frac{A_2^2N_1N_2(g_{12}-g)}{\sqrt{2 \pi } \sigma _2},\\
        E_{\rm LHY} &=- \frac{(2 g N_{\rm w})^{3/2}}{3\sqrt{3} \pi^{5/4}\sqrt{\sigma_1}}
         - \frac{(g \tilde{N})^{3/2} \mathcal{R}(-\Lambda)}{\sqrt{6} \pi^{5/4}\sqrt{\sigma_2}}.
    \end{align}
\end{subequations}}
In Eqs.~(\ref{E_comp_sm}a)-(\ref{E_comp_sm}c), $\tilde{N} =\int_{-\infty}^{\infty} dx \left(|\phi_1^{\rm in}|^2+|\phi_2|^2\right)=A_2^2 N_1 + N_2$ and $N_{\rm w} =\int_{-\infty}^{\infty} dx |\phi_1^{\rm out}|^2= (1 - A_2^2) N_1$ are the variational estimates of effective total particle numbers in the droplet core and its outside wings, respectively, and 
$\Lambda = 4 A_2^2 N_1 N_2 \eta^2 / \tilde{N}^2$ and $\eta = \sqrt{1-\gamma^2}$ (see the Supplemental Material \cite{SupplementalMaterial}). The Euler-Lagrange equations for $\sigma_1(t)$ and $\sigma_2(t)$ thus obtained are
\begin{subequations}
\begin{align}
    \ddot{\sigma}_1 &= \frac{1}{\sigma_1^3} + \frac{ g N_{\rm w}}{\sqrt{2 \pi } \sigma_1^2} - \frac{2 \sqrt{2} g^{\!3/2} \sqrt{N_{\rm w}}}{3 \sqrt{3}\pi ^{5/4} \sigma_1^{3/2}}, \\
    \ddot{\sigma}_2 &= \frac{1}{\sigma_2^3} +\frac{g\tilde{N}}{\sqrt{2 \pi }\sigma _2^2}+ \frac{\sqrt{2} A_2^2 N_1 N_2(g_{12}-g)}{\sqrt{\pi }\tilde{N} \sigma _2^2} \nonumber\\
                    &- \frac{g^{3/2} \sqrt{\tilde{N}}\mathcal{R}(-\Lambda)}{\sqrt{6}\pi^{5/4}\sigma_2^{\!3/2}}\label{sigma2},
\end{align}
\end{subequations}
Linearization of
the Euler-Lagrange equations yields $\delta\ddot\sigma_\nu(t) + \omega_\nu^2\delta\sigma_\nu(t)=0$ with
\begin{subequations}
\begin{align}
    \omega_1^2 &= \frac{3}{\bar\sigma_1^4} + \sqrt{\frac{2}{\pi}} \frac{g N_{\rm w}}{\bar \sigma_1^3}- \sqrt{\frac{2}{3}} \frac{g^{\!3/2} \sqrt{N_{\rm w}}}{\pi ^{5/4} \bar\sigma_1^{5/2}},\label{omega1} \\
    \omega_2^2 &= \frac{3}{\bar\sigma_2^4} + \sqrt{\frac{2}{\pi}} \frac{g\tilde{N}}{\bar\sigma_2^3}+ \frac{2\sqrt{2} A_2^2N_1 N_2(g_{12}-g)}{\sqrt{\pi}\tilde{N}\bar\sigma_2^3} \nonumber\\&- \frac{\sqrt{3 \tilde{N}} g^{3/2}\mathcal{R}(-\Lambda)}{2\sqrt{2}\pi^{5/4}\bar\sigma_2^{5/2}}\label{omega2},
\end{align}
\label{omega12}
\end{subequations}
where $\bar \sigma_1$ and $\bar \sigma_2$ are the equilibrium measures of widths.% \sout{$\sigma_1$ and $\sigma_2$ respectively}}. 
These, along with $A_2$, are fixed by minimizing the ground-state variational energy
\begin{equation}
\begin{split}
E_0 &= \frac{N_{\rm w}}{4\sigma_1^2} + \frac{\tilde{N}}{4\sigma_2^2}
  + \frac{g N_{\rm w}^2}{2\sqrt{2\pi} \sigma_1}+ \frac{A_2^2 N_1 N_2(g_{12}-g)}{\sqrt{2 \pi } \sigma _2}
  \\&  + \frac{g\tilde{N}^2}{2 \sqrt{2 \pi } \sigma _2} - \frac{(2 g N_{\rm w})^{3/2}}{3\sqrt{3} \pi^{5/4}\sqrt{\sigma_1}} - \frac{(g \tilde{N})^{3/2} \mathcal{R}(-\Lambda)}{\sqrt{6} \pi^{5/4}\sqrt{\sigma_2}}.\label{E_var}
\end{split}
\end{equation}
Equations (\ref{omega12}) and (\ref{E_var}) describe the variational estimates for $\delta N\ge8$.
For $\delta N<8$, all the atoms of the majority component are bound within the droplet. In this case, we set $A_1 = 0$ and $A_2= 1$, so the Lagrangian's dependence
on $\phi_1^{\rm out}$ [or $\sigma_1(t), \beta_1(t)$ and their derivatives] drops out. The equation of motion of the droplet's size $\sigma_2(t)$
is again given by Eq.~(\ref{sigma2}) and density-breathing mode frequency by $\omega_2$ in Eq.~(\ref{omega2}). The variational results for $\omega_1$ and
$\omega_2$ are shown in Fig.~\ref{exc_2com}(b) and are in good agreement with the corresponding modes in the BdG spectrum.
The variational analysis thus clearly demonstrates the origin of the beat pattern in the breathing dynamics of the majority component, 
which is characterized by two length scales, namely $\sigma_1$ and $\sigma_2$.

\section{Self-bound Droplet to Unbound Gas Transition}
\label{Sec-IV}
\begin{figure}[!htbp]
\centering
\includegraphics[width=\columnwidth]{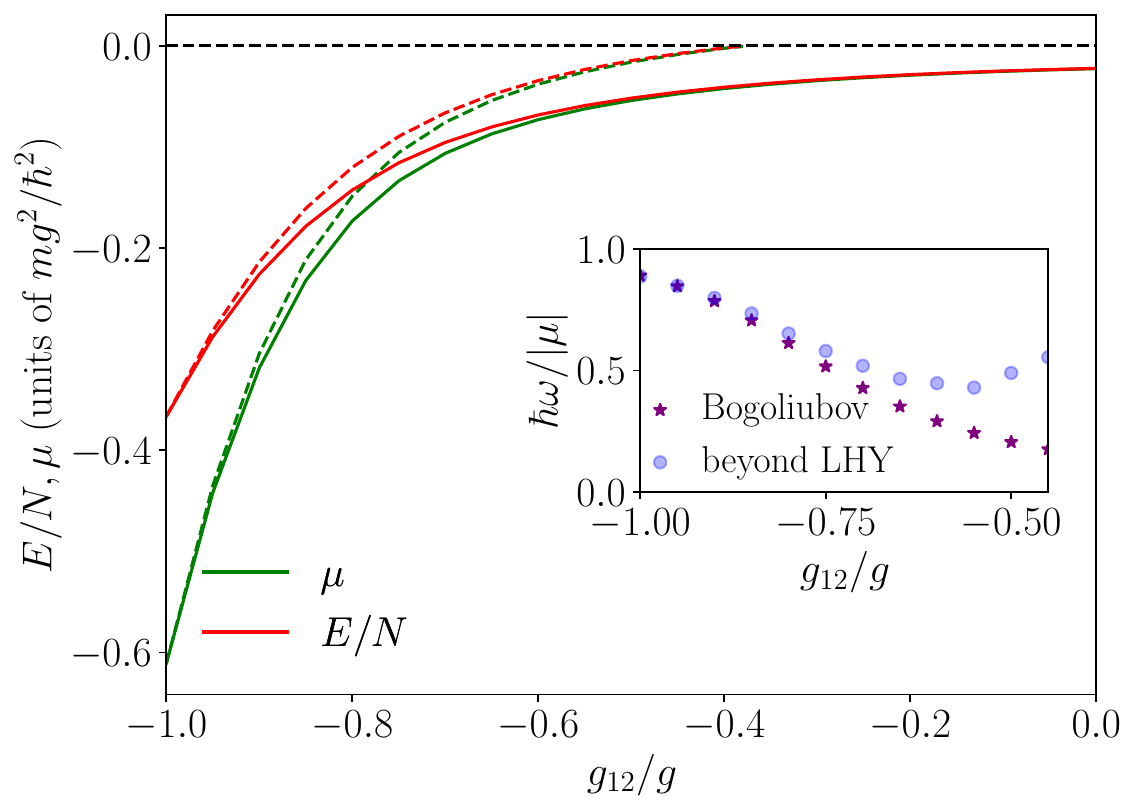}
\caption{Comparison of the energy per particle and chemical potential obtained from the Bogoliubov theory (solid lines) and the beyond-LHY theory (dashed lines) for a pseudospinor mixture with $N=20$. The inset shows the energy of the density-breathing mode, scaled by the chemical potential, for these two theories.}
\label{energy}
\end{figure}
The Bogoliubov theory supports the droplet phase across the entire mean-field stability regime \cite{10.21468/SciPostPhys.19.5.133}. However, the weak-coupling limit $n|a_{\nu\nu'}|\gg1$ degrades as
%\sout{The validity of the theory requires that $n|a_{\nu\nu'}|\gg1$ and since the} 
the peak density decreases with an increase in $\delta g$ before
getting saturated. Hence, the Bogoliubov theory's validity 
is not guaranteed across the entire mean-field stability regime. Furthermore, QMC studies have shown that there is a liquid-to-unbound-gas
transition for $g_{12}>-0.46g$ in a pseudospinor 1D Bose-Bose mixture~\cite{ PhysRevLett.122.105302, PhysRevA.102.023318}, in contrast
to the prediction of the Bogoliubov theory \cite{10.21468/SciPostPhys.9.2.020,10.21468/SciPostPhys.19.5.133}. 

In this section we implement a beyond-LHY description of Bose-Bose mixtures \cite{10.21468/SciPostPhys.9.2.020}, using beyond mean-field expressions for sound velocities 
in Eq. (\ref{mu_lhy}), to validate the regimes studied in this work, and to locate the droplet-to-gas transition and to reexamine the density-breathing mode independently. 
We again consider $g_{11} = g_{22} = g$, which sets the equilibrium density difference to zero for a pseudospinor mixture, but not for a scalar mixture
with population imbalance ($\delta N>0$) [see Fig.~\ref{den_2com}(f)].
To obtain the beyond-mean-field expressions for sound velocities applicable for both these systems,  
we first calculate the compressibility $\kappa_-$ and susceptibility $\kappa_+$ using \cite{10.21468/SciPostPhys.9.2.020}
\begin{equation}
    \kappa_- = \frac{1}{n^2}\left(\frac{\partial^2 \mathcal{E}}{\partial n^2}\right)^{-1}, \kappa_+ = \frac{1}{n^2}\left(\frac{\partial^2 \mathcal{E}}{\partial \left(\Delta n\right)^2}\right)^{-1},
\end{equation}
where $\mathcal{E}$ is the energy density (\ref{ener}), $n = n_1+n_2$ and $\Delta n = n_1-n_2$. Then using $c_{\pm}=\sqrt{1/(n\kappa_{\pm})}$, we obtain the beyond mean-field expressions for sound velocities and their respective derivatives
\begin{subequations}\label{vel}
\begin{align}
c_+^2 &=  \frac{n(g-g_{12})}{2} - \frac{g\eta^2\sqrt{2gmn}}{8\hbar\pi}\nonumber\\& \quad \left(\frac{\Delta n^2 \eta^2 \mathcal{Q}}{n^2(p+1)}-4\sqrt{2}\left(\frac{\Delta n^2\eta^2}{n^2(p+1)}-1\right)\frac{\partial \mathcal{R}}{\partial p}\right),\\
c_-^2 &= \frac{n(g+g_{12})}{2}-\frac{g\sqrt{2gmn}}{8\hbar\pi(p+1)}\nonumber\\& \quad \left((1+p+\gamma^4)\mathcal{Q} + 4\sqrt{2}\gamma^2 \left(\frac{\Delta n^2 \eta^2}{n^2}-\frac{p+1}{\sqrt{-p}}\right)\frac{\partial \mathcal{R}}{\partial p}\right),
\end{align}
\end{subequations}

\begin{widetext}
\begin{subequations}\label{dvel}
\begin{align}
\frac{\partial c_+^2}{\partial n_\nu} &= \frac{(g-g_{12})}{2}-\frac{g\eta^2\sqrt{2gmn}}{8\hbar\pi}\Bigg[\frac{\Delta n^2\eta^2\mathcal{Q}}{2n^3(p+1)}+\frac{2\mathcal{Q}\eta^2\Delta n}{n^2(p+1)}\left(s_\nu-\frac{\Delta n}{n}\right) +\frac{\Delta n^2\eta^2}{n^2(p+1)}\frac{\partial p}{\partial n_\nu}\left(\frac{\partial \mathcal{Q}}{\partial p}-\frac{\mathcal{Q}}{1+p}\right)\nonumber \\& \quad -\frac{4\sqrt{2}\Delta n\eta^2}{n^2(p+1)}\frac{\partial \mathcal{R}}{\partial p}\left(2(-1)^{\nu+1}-\frac{2\Delta n}{n}-\frac{\Delta n}{p+1}\frac{\partial p}{\partial n_\nu}\right)-2\sqrt{2}\left(\frac{\Delta n^2\eta^2}{n^2(p+1)}-1\right)\left(\frac{1}{n}\frac{\partial \mathcal{R}}{\partial p}+2\frac{\partial^2 \mathcal{R}}{\partial p^2}\frac{\partial p}{\partial n_\nu}\right)  \Bigg], \\
\frac{\partial c_-^2}{\partial n_\nu} &= \frac{(g+g_{12})}{2}-\frac{g\sqrt{2gmn}}{8\hbar\pi(p+1)}\Bigg[\left(\frac{1}{2n}-\frac{1}{p+1}\frac{\partial p}{\partial n_\nu}\right)\left(1+p+\gamma^4\right)\mathcal{Q}+\left(\mathcal{Q}+\left(1+p+\gamma^4\right)\frac{\partial \mathcal{Q}}{\partial p}\right)\frac{\partial p}{\partial n_\nu}\nonumber \\& \quad + 4\sqrt{2}\gamma^2\left(\frac{\Delta n^2 \eta^2}{n^2}-\frac{p+1}{\sqrt{-p}}\right)\left(\frac{\partial^2 \mathcal{R}}{\partial p^2}\frac{\partial p}{\partial n_\nu}+\frac{\partial \mathcal{R}}{\partial p}\left(\frac{1}{2n}-\frac{1}{p+1}\frac{\partial p}{\partial n_\nu}\right)\right)\nonumber \\& \quad+4\sqrt{2}\gamma^2\frac{\partial \mathcal{R}}{\partial p}\left(\frac{2\Delta n \eta^2}{n^2}\left(s_\nu-\frac{\Delta n}{n}\right)-\frac{1}{\sqrt{-p}}\frac{\partial p}{\partial n_\nu}\left(1-\frac{p+1}{2p}\right)\right)\Bigg]
\end{align}
\end{subequations}
\end{widetext}
where $s_\nu = (-1)^{\nu+1}$,
\begin{subequations}
    \begin{align}
    p &= \frac{4(g_{12}^2-g^2)n_1n_2}{g^2n^2},\\
    \mathcal{Q}(p) &= \frac{1}{\sqrt{1-\sqrt{p+1}}}+\frac{1}{\sqrt{1+\sqrt{p+1}}}, 
\end{align}
\end{subequations}
and $\mathcal{R}(p)$ is defined in Eq.~(\ref{Rp}). 
The beyond-mean-field sound velocities $c_{\pm}$ in Eqs.~(\ref{vel}a) and (\ref{vel}b) are substituted in Eq.~(\ref{ener}) to
obtain the beyond-Bogoliubov expression for the energy density. Similarly, Eq.~(\ref{mu_lhy}) along with Eqs.~(\ref{vel}) and  (\ref{dvel}), provides the beyond-LHY correction term appearing in the GP equation~(\ref{eeGPE}). We emphasize that the beyond-LHY description modifies the quantum-fluctuation-generated correction itself, via the beyond-mean-field sound velocities obtained from the compressibility and susceptibility through Eqs.~(\ref{vel}) and  (\ref{dvel}), rather than operating at the mean-field level.

\begin{figure}[!htbp]
\centering
\includegraphics[width=\columnwidth]{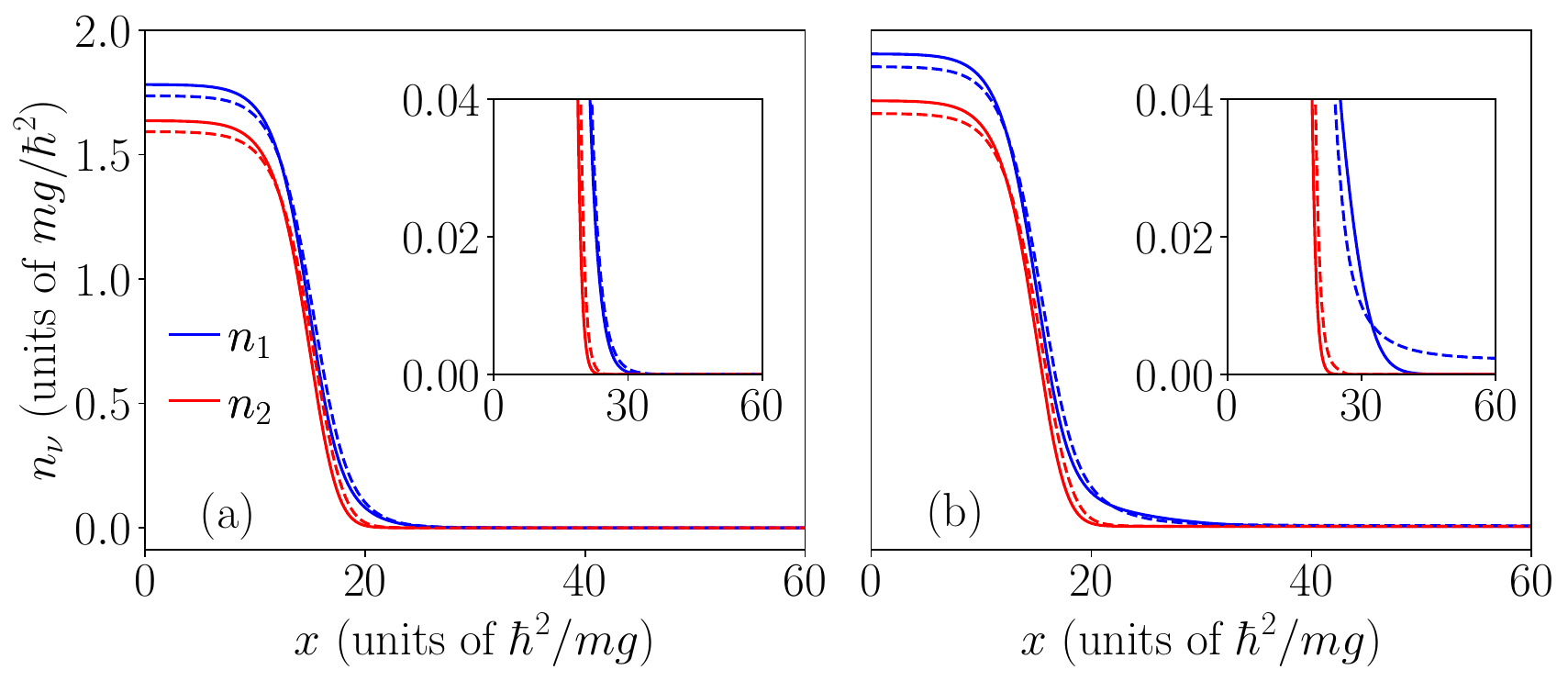}
\caption{Density profiles comparing the Bogoliubov theory (solid lines) and the beyond-LHY theory (dashed lines) for $N=100$ and $g_{12}=-0.8g$ at two different values of population imbalance $\delta N$: (a) $\delta N = 6$ and (b) $\delta N = 8$. The insets provide a closer view of the behavior near the edges of the droplet. At $\delta N = 6$, both models exhibit vanishing densities near the edges. However, at $\delta N = 8$, the Bogoliubov model still predicts vanishing densities for both components near the edges; the beyond-LHY model gives a finite edge density for the majority component, which ends up occupying the
entire 1D box size considered in the simulation, and zero density for the minority component. As the densities are symmetric about $x=0$, i.e. $n_\nu(x) = n_{\nu}(-x)$, densities are shown only for $+x$ values.}
\label{den_2com_blhy}
\end{figure}
The beyond-LHY description results in a liquid-to-gas transition at $g_{12}\approx-0.38g$ for a pseudospinor mixture \cite{10.21468/SciPostPhys.9.2.020}. This is illustrated in Fig.~\ref{energy}, where the ground-state energy per particle and the chemical potential are plotted as a function of $g_{12}/g$ for $N = 20$ using the Bogoliubov theory and the beyond-LHY description. We can see that at $g_{12}\approx-0.38g$, $E/N$ and $\mu$ calculated using the beyond-LHY approach are $0$; after that, the droplet is no longer self-bound. However, in the Bogoliubov theory, $E/N$ and $\mu$ are negative even for $g_{12}>-0.38g$ and the droplet is still self-bound \cite{10.21468/SciPostPhys.9.2.020}. 
We calculate the density-breathing mode from the real-time evolution of the ground state within the beyond-LHY prescription, i.e., using Eq.~(\ref{eeGPE}) with Eq.~(\ref{mu_lhy}) evaluated
using the beyond-mean-field estimates of phonon velocities and their derivatives.
The inset in Fig.~\ref{energy} shows the comparison of the density-breathing mode's energies obtained from the Bogoliubov theory and the beyond-LHY prescription. 
Within the beyond-LHY formalism, the density-breathing mode's energy scaled with respect to the particle-emission threshold first decreases with an increase in $g_{12}/g$,
and near the liquid-to-gas transition point, it starts increasing as the particle-emission threshold approaches zero. 

\begin{figure}[!htbp]
\centering
\includegraphics[width=0.85\columnwidth]{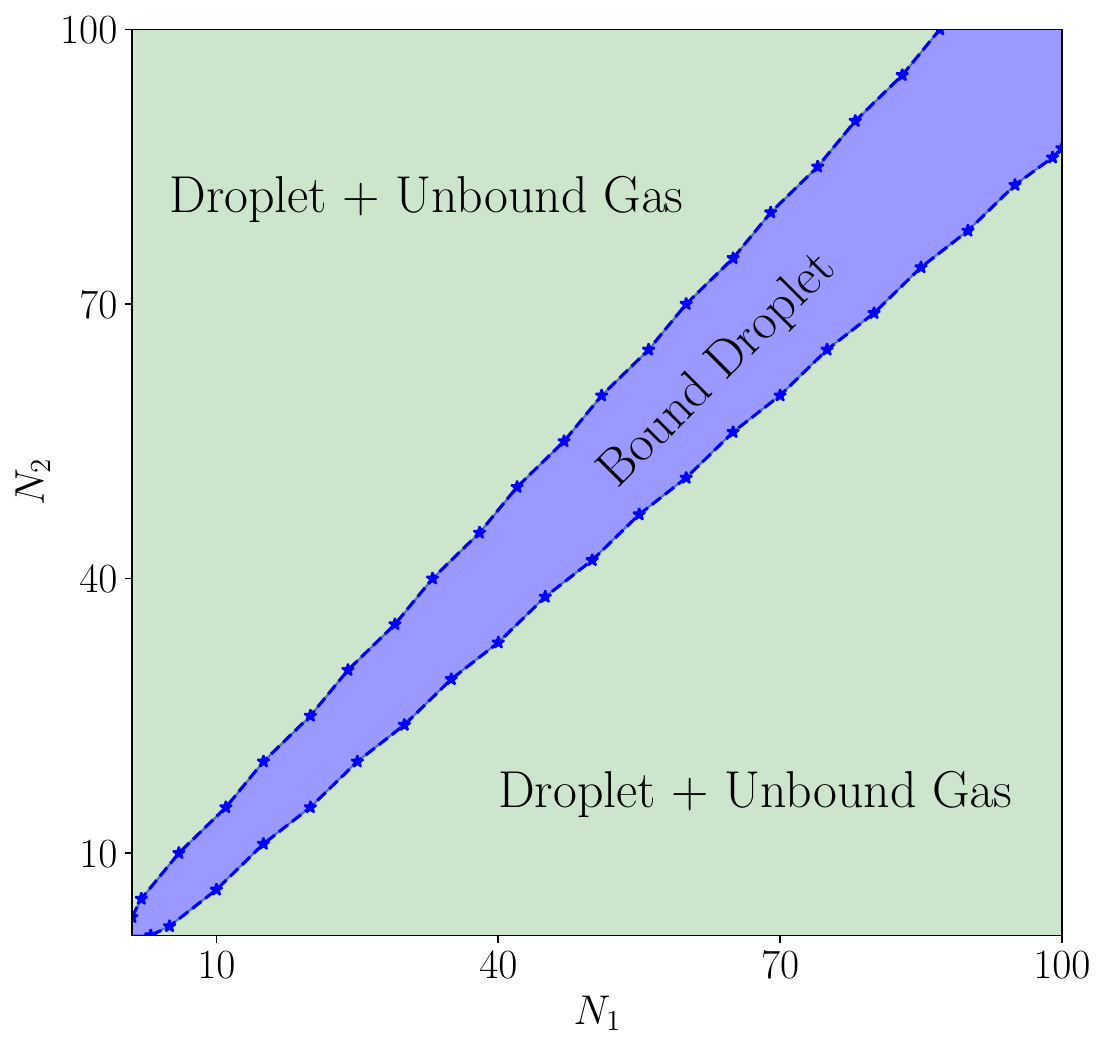}
\caption{Phase diagram obtained using the beyond-LHY model in the $N_1$-$N_2$ plane for $g_{12}=-0.8g$. The blue shaded region, labeled ``Bound Droplet," represents the regime where both components are bound and have negative chemical potentials. The green shaded region, labeled ``Droplet+Unbound Gas," represents the regime where the self-bound droplet coexists with an unbound gas of the majority component's atoms. In this region, the chemical potential of the majority component is zero, whereas that of the minority component is still negative.}
\label{phase}
\end{figure}

We also use the beyond-LHY description for the scalar mixture discussed in Sec. \ref{Sec-III}. The beyond-LHY chemical potentials 
of the two components are plotted in Fig.~\ref{exc_2com}(a). The chemical potential of the majority component approaches zero with an increase of particle imbalance, 
and for $\delta N \geq 8$, $\mu_1\approx 0$, and we have an unbound gas of the majority component outside the droplet. This is further illustrated in Figs. \ref{den_2com_blhy}(a)
and \ref{den_2com_blhy}(b), showing a comparison of the density profiles of the solutions obtained with the Bogoliubov and beyond-LHY theories. For $\delta N <8$, both theories result in 
localized self-trapped solutions with vanishing densities outside the droplet, as can be seen in the inset of Fig. \ref{den_2com_blhy}(a) (for $\delta N=6$). For $\delta N= 8$, 
beyond-LHY theory results in finite density of the majority component outside the droplet filling up the entire extent of the simulation box, however large,  unlike the density profile obtained from the Bogoliubov theory, which is still confined within the simulation box [see inset of Fig. \ref{den_2com_blhy}(b)]. Additionally, we examined
the density-breathing mode of the system from the real-time evolution of the ground state for $\delta N<8$ within the beyond-LHY prescription, namely Eq.~(\ref{eeGPE}) with Eq.~(\ref{mu_lhy}) evaluated using the beyond-mean-field estimates of phonon velocities and their derivatives. The energies of the breathing excitation as a function of 
$\delta N$ thus obtained are slightly lower than the corresponding numbers obtained from the Bogoliubov theory as shown in Fig.~\ref{exc_2com}(a).
 The phase diagram of the scalar mixture using the beyond-LHY description in the $N_1$-$N_2$ plane for $g_{12} = -0.8g$ 
is shown in Fig.~\ref{phase}. The bound droplet phase exists for a finite range of population imbalances around diagonal $N_1=N_2$ line, and as one moves along it, the phase's domain widens (see blue
shaded region in Fig. \ref{phase}). The region outside this wedge-shaped region, the light green region in Fig. \ref{phase}, corresponds to the droplet inside an unbound gas of 
the majority species' atoms.

\section{Summary and conclusions}
\label{summary}
We have conducted theoretical and numerical studies of the collective excitations of one-dimensional quantum droplets
in pseudospinor and scalar Bose-Bose mixtures. We employed Bogoliubov theory to access both density (in-phase) and
spin (out-of-phase) sectors and computed the excitation spectrum from the numerical solutions of the Bogoliubov-de Gennes
equations. Our findings indicate that for droplets realized within the mean-field stability regime, spin modes can
soften and fall below the particle-emission threshold, making them observable in the droplet spectrum.

To complement the full BdG calculations, we developed a variational description of the density- and spin-breathing modes.
The variational description of the density- and spin-breathing modes not only agrees well with the BdG spectrum, providing an independent check on the numerics, but also yields a transparent physical picture of the modes as coupled width oscillations of the two components. Finally, we corroborated the mode
analysis by studying the real-time dynamics of the breathing mode. 

For imbalanced droplets, we demonstrated that as the imbalance increases above a critical number, %($\delta N \geq 8$), 
two distinct length scales emerge. One is associated with the overlap or core region of the self-bound droplet, while the other corresponds to the halo of the unbound atoms of the majority component. We explained how these two length scales govern the droplet's breathing dynamics, demonstrating that the mode frequencies calculated via the BdG formalism are in excellent agreement with both real-time simulations and variational estimates.
We complemented these studies with an investigation of the liquid-to-gas transition and the density-breathing mode, in both pseudospinor
and population-imbalanced scalar mixtures, using a beyond-LHY theory.

 In contrast to three-dimensional droplets~\cite {doi:10.1126/science.aao5686, PhysRevLett.120.235301, PhysRevLett.120.135301, PhysRevLett.122.090401, PhysRevResearch.1.033155, Wang_PhysRevResearch.3.033247, Burchianti_PhysRevLett.134.093401, Kadau2016, PhysRevLett.116.215301, Schmitt2016, PhysRevX.6.041039}, one-dimensional Bose-Bose droplets have not yet been realized experimentally, though the phase is established by QMC~\cite{ PhysRevLett.122.105302, PhysRevA.102.023318}. Nonetheless, the required quasi-1D confinement 
and Feshbach tunability of the couplings are available \cite{PhysRevLett.120.135301}, making the realization of such droplets, which
are expected to exhibit reduced three-body recombination losses \cite{PhysRevLett.92.190401, PhysRevA.103.043316} and enhanced beyond mean-field effects, experimentally feasible in the future.
The density-breathing mode is accessible via a coupling quench and rms-width tracking; the spin-breathing mode, being an out-of-phase oscillation, is addressable through spin-resolved
imaging and out-of-phase driving. Since the particle-emission threshold sets which modes are bound, the softening of spin modes below it constitutes an experimentally accessible signature.

{\em Note added.} Recently, we noticed a recent paper by Xiao {\em et al.} \cite{2zbp-j5bg} which investigates ground-state properties and collective excitations of one-dimensional asymmetric quantum droplets. They specifically evaluate spin-dependent collective modes under confinement, providing an interesting parallel to our results for the spin modes of the self-trapped state presented here.
\section*{ACKNOWLEDGMENT}
The work of R.K.G. was supported by SERB Grant No. CRG/2023/001388.
\section*{DATA AVAILABILITY}
There are no publicly available research data or software
supporting this paper. Requests for further information or data
should be sent to the authors.
%\section*{Appendix: }
\begin{widetext}
\begin{center}
\textbf{APPENDIX A: BdG matrix elements}
\end{center}
The elements of the BdG matrix in Eq.~(\ref{bdg}) are 
\begin{subequations}
    \begin{align}
        \mathcal{M} &= -\frac{\hbar^2}{2m}\frac{\partial^2}{\partial x^2} + 2g_{11}|\phi_1|^2+g_{12}|\phi_2|^2 -\mu_1 + A_{11}, \\ \mathcal{N} &= -\frac{\hbar^2}{2m}\frac{\partial^2}{\partial x^2} + 2g_{22}|\phi_2|^2+g_{21}|\phi_1|^2 -\mu_2 + A_{22},
    \end{align}
\label{matrix_element}
\end{subequations}
where 
\begin{subequations}
    \begin{align}
        A_{\nu'\nu} &= -\frac{g_{\nu\nu}\sqrt{m(g_{11}n_1+g_{22}n_2)}}{\hbar \pi}\left(\frac{g_{\nu'\nu'}\phi_\nu\phi_{\nu'}^*}{2(g_{11}n_1+g_{22}n_2)}+\frac{\partial \phi_\nu}{\partial \phi_{\nu'}}\right), \\ B_{\nu'\nu}&= -\frac{g_{\nu\nu}\sqrt{m}}{\hbar \pi}\frac{g_{\nu'\nu'}\phi_\nu\phi_{\nu'}}{2(g_{11}n_1+g_{22}n_2)},
    \end{align}
\label{matrix_element_petrov}
\end{subequations}
for Petrov's model and
%\begin{widetext}
\begin{subequations}
    \begin{align}
        A_{\nu'\nu} &= -\frac{3g_{\nu\nu}\sqrt{m(g_{11}n_1+g_{22}n_2)}}{4\hbar\pi}\Bigg(\frac{g_{\nu'\nu'}\mathcal{R}\phi_{\nu'}^*\phi_\nu}{2(g_{11}n_1+g_{22}n_2)}+\phi_{\nu'}^*\phi_\nu\frac{\partial\mathcal{R}}{\partial p}\frac{\partial p}{\partial n_{\nu'}}+\mathcal{R}\frac{\partial \phi_\nu}{\partial \phi_{\nu'}}+\frac{g _{\nu'\nu'}}{g_{\nu\nu}}\phi_{\nu'}^*\phi_\nu\frac{\partial\mathcal{R}}{\partial p}\frac{\partial p}{\partial n_\nu}\Bigg)\nonumber \\&- \frac{\sqrt{m(g_{11}n_1+g_{22}n_2)^3}}{2\hbar\pi}\Bigg(\phi_{\nu'}^*\phi_\nu\frac{\partial^2\mathcal{R}}{\partial p^2}\frac{\partial p}{\partial n_{\nu'}}\frac{\partial p}{\partial n_\nu}+\frac{\partial\mathcal{R}}{\partial p}\frac{\partial^2 p}{\partial n_\nu'\partial n_\nu}\phi_{\nu'}^*\phi_\nu+\frac{\partial\mathcal{R}}{\partial p}\frac{\partial p}{\partial n_\nu}\frac{\partial \phi_\nu}{\partial \phi_{\nu'}}\Bigg),\\
        B_{\nu'\nu} &= -\frac{3g_{\nu\nu}\sqrt{m(g_{11}n_1+g_{22}n_2)}}{4\hbar\pi}\Bigg(\frac{g_{\nu'\nu'}\mathcal{R}\phi_{\nu'}\phi_\nu}{2(g_{11}n_1+g_{22}n_2)}+\phi_{\nu'}\phi_\nu\frac{\partial\mathcal{R}}{\partial p}\frac{\partial p}{\partial n_{\nu'}}+ \frac{g _{\nu'\nu'}}{g_{\nu\nu}}\phi_{\nu'}\phi_\nu\frac{\partial\mathcal{R}}{\partial p}\frac{\partial p}{\partial n_\nu}\Bigg)\nonumber\\& - \frac{\sqrt{m(g_{11}n_1+g_{22}n_2)^3}}{2\hbar\pi}\Bigg(\phi_{\nu'}\phi_\nu\frac{\partial^2\mathcal{R}}{\partial p^2}\frac{\partial p}{\partial n_{\nu'}}\frac{\partial p}{\partial n_\nu}+\frac{\partial\mathcal{R}}{\partial p}\frac{\partial^2 p}{\partial n_{\nu'}\partial n_\nu}\phi_{\nu'}\phi_\nu\Bigg),
    \end{align}
\label{matrix_element_bogo}
\end{subequations}
for the Bogoliubov model.
%\end{widetext}
In Eqs.~(\ref{matrix_element_bogo}a) and (\ref{matrix_element_bogo}b)
\begin{subequations}
    \begin{align}
        \frac{\partial^2p}{\partial n_\nu^2} &= \frac{4g_{\nu\nu}n_{\nu'}(g_{11}g_{22}-g_{12}^2)}{(g_{11}n_1+g_{22}n_2)^3}-\frac{3g_{\nu\nu}}{(g_{11}n_1+g_{22}n_2)}\frac{\partial p}{\partial n_\nu}, \\
        \frac{\partial^2p}{\partial n_{\nu'} \partial n_\nu} &= \frac{4(g_{11}g_{22}-g_{12}^2)(g_{\nu\nu}n_\nu-2g_{\nu'\nu'}n_{\nu'})}{(g_{11}n_1+g_{22}n_2)^3} -\frac{3g_{\nu'\nu'}}{(g_{11}n_1+g_{22}n_2)}\frac{\partial p}{\partial n_\nu}
    \end{align}
\end{subequations}
\begin{center}

\textbf{%\sout{Coefficients in $L_{\rm LHY}$} 
APPENDIX B: $E_{\rm LHY}$ and the Euler-Lagrange equation for $\sigma_{\nu}$ of the pseudospinor system }

\end{center}
We rewrite Eq.~(\ref{L_LHY_int}) by expressing the integrand as a function of $x,\bar\sigma(t),$ and $\epsilon(t)$ and then
employ a single-variable Taylor series expansion in $\epsilon(t)$ about $\epsilon(t) = 0$ to obtain 
\begin{align}
E_{\rm LHY} &= \int f(x, \bar \sigma,\epsilon) dx,\label{elhy_int}\\
            & = \int \left[f(x, \bar\sigma,0)  + \frac{1}{2}\frac{\partial^2 f}{\partial\epsilon^2}\Bigg|_{\epsilon=0}\epsilon^2 +O(\epsilon^4)\right] dx,
\end{align}
where 
\begin{align}
f(x, \bar \sigma,\epsilon)&= -\frac{\left(g N \right)^{3/2}}{12 \pi ^{7/4}}\left(\frac{e^{-x^2/(\bar\sigma +\epsilon)^2}}{\bar\sigma +\epsilon}+\frac{e^{-x^2/(\bar\sigma -\epsilon )^2}}{\bar\sigma -\epsilon}\right)^{3/2}\left[\left(1-\sqrt{1-\frac{4 \left(1-\gamma^2\right) (\bar\sigma^2 -\epsilon ^2) e^{2x^2(\bar\sigma^2+\epsilon^2)/(\bar\sigma^2-\epsilon^2)^2}}{ \left[(\bar\sigma -\epsilon) e^{x^2/(\bar\sigma -\epsilon )^2}+(\bar\sigma +\epsilon) e^{x^2/(\bar\sigma +\epsilon )^2}\right]^2}}\right)^{3/2}\right.\nonumber\\&\left.+\left(1+\sqrt{1-\frac{4 \left(1-\gamma^2\right) (\bar\sigma^2 -\epsilon ^2)  e^{2x^2(\bar\sigma^2+\epsilon^2)/(\bar\sigma^2-\epsilon^2)^2}}{\left[(\bar\sigma -\epsilon ) e^{x^2/(\bar\sigma -\epsilon )^2}+(\bar\sigma +\epsilon) e^{x^2/(\bar\sigma +\epsilon )^2}\right]^2}}\right)^{3/2}\right]
\end{align}
\begin{subequations}\label{d2fdeps2}
    \begin{align}
f(x, \bar \sigma,0)&=\frac{C_1 }{3 \sqrt{2} \pi ^{7/4}}\left(\frac{g N e^{-\frac{x^2}{\bar \sigma ^2}}}{\bar \sigma }\right)^{3/2}, \\\frac{\partial^2 f}{\partial\epsilon^2}\Bigg|_{\epsilon=0}&= \frac{\mathcal{C}_2 \left(\bar \sigma^2-2 x^2\right)^2+2 \mathcal{C}_1 \left(\bar \sigma^4-5 \bar \sigma^2 x^2+2 x^4\right)}{2 \sqrt{2} \pi ^{7/4}\bar \sigma^6}\left(\frac{g N e^{-\frac{x^2}{\bar \sigma^2}}}{\bar \sigma}\right)^{3/2}. 
\end{align}
\end{subequations}
In Eqs. (\ref{d2fdeps2}a) and  (\ref{d2fdeps2}b),
 \begin{subequations}    \label{c1c2}
  \begin{align}
       \mathcal{C}_1 &= -\Big[(1-|\gamma|)^{3/2}+(1+|\gamma|)^{3/2}\Big],\\
      \quad \mathcal{C}_2 &= \frac{(1-\gamma^{2})\big(\sqrt{1-|\gamma|}-\sqrt{1+|\gamma|}\big)}{|\gamma|},
\end{align}  
\end{subequations}
where $\gamma = g_{12}/g$. For Petrov's model, $|\gamma|=1$, $\mathcal{C}_1 = -2^{3/2}$ and $\mathcal{C}_2=0$.
As the integrand in (\ref{elhy_int}) is an even function of $\epsilon$,  reflecting the symmetry with respect to the mutual interchange of $\sigma_1(t)$ and $\sigma_2(t)$, only even powers of $\epsilon$ appear
in the expansion. We truncate the series up to the second order in $\epsilon(t)$ to obtain
\begin{align}
E_{\rm LHY} &\approx \frac{2 (g N)^{3/2}}{3\sqrt{6}\pi ^{5/4} (\sigma_1+\sigma_2)^{5/2}}\Bigg[\mathcal{C}_1 (\sigma_1+\sigma_2)^2+\frac{3}{4} \mathcal{C}_2  (\sigma_1-\sigma_2)^2\Bigg].\label{E_LHY}
\end{align}
Equation~(\ref{lagrangian}) with $E_{\rm MF}$  [Eq.~(\ref{E_MF})] and $E_{\rm LHY}$ [Eq.~(\ref{E_LHY})] defines the variational Lagrangian of the system and is used to derive the Euler-Lagrange equations \cite{SupplementalMaterial}.
The Euler-Lagrange equation for the variational parameter $\sigma_\nu$ as referred to in the discussion on the variational method for the pseudospinor mixture in Sec.~\ref {Sec-II} is
\begin{equation}
\begin{split}
\ddot \sigma_\nu&= \frac{1}{\sigma_\nu^3}+\frac{N}{ \sqrt{\pi } \sigma_\nu^2}\left(\frac{g}{2\sqrt{2}}+\frac{g_{12}\sigma_\nu^3}{\left(\sigma_{\bar\nu}^2+\sigma_\nu^2\right)^{3/2}}\right)+\frac{g^{3/2}\sqrt{N}}{\sqrt{6}\pi^{5/4}(\sigma_{\bar\nu}+\sigma_\nu)^{7/2}}\Bigg[\frac{4}{3}\mathcal{C}_1(\sigma_{\bar\nu}+\sigma_\nu)^{2}+\mathcal{C}_2(\sigma_\nu-\sigma_{\bar\nu})(\sigma_\nu-9\sigma_{\bar\nu})\Bigg],\\
                &= g_\nu(\sigma_1,\sigma_2).
\end{split}
\label{condensate_width}
\end{equation}
The linearization of the equation yields
\begin{align}
\delta \ddot{\sigma}_\nu &= \frac{\partial g_\nu}{\partial \sigma_\nu}\Bigg|_{\sigma_\nu=\sigma}\delta \sigma_\nu+\frac{\partial g_\nu}{\partial \sigma_{\bar \nu}}\Bigg|_{\sigma_{\bar \nu}=\sigma}\delta \sigma_{\bar \nu},\\ 
                         &= -\omega_{\nu\nu}^2\delta \sigma_\nu- \omega_{\nu\bar{\nu}}^2\delta \sigma_{\bar\nu},
\end{align}
where $\nu$ and $\bar\nu$ each take values $1, 2$ with $\nu\ne\bar\nu$, and $\omega_{\nu\nu}^2=\omega_1^2$ and $\omega_{\nu\bar{\nu}}^2 =\omega_2^2$ are defined in Eqs. (\ref{omega11}) and (\ref{omega22}), respectively.

The
Taylor series expansion up to second order in $\epsilon(t) = [\sigma_1(t)-\sigma_2(t)]/2$  is sufficient to approximate the 
integrand in (\ref{L_LHY_int}). 
To explicitly show this, we express Lagrangian for the pseudospinor system as
\begin{align}
    L = -\frac{N}{4}\left(\sigma_1^2\dot \beta_1+\sigma_2^2\dot \beta_2+2\sigma_1^2 \beta_1^2+2\sigma_2^2 \beta_2^2\right) - V_{\rm eff}(\sigma_1,\sigma_2),
\end{align}
where 
\begin{align}
    V_{\rm eff}(\sigma_1,\sigma_2) = \frac{N}{8}\left(\frac{1}{\sigma_1^2}+\frac{1}{\sigma_2^2}\right)+E_{\rm MF} + E_{\rm LHY}
\end{align}
The Euler-Lagrange equation for $\sigma_\nu$ then 
gives
\begin{align}
    -\frac{N\sigma_\nu}{2}\left(\dot\beta_\nu+2\beta_\nu^2\right) - \frac{\partial V_{\rm eff}}{\partial \sigma_\nu} = 0,
\end{align}
which on substituting $\beta_\nu =  \dot{\sigma}_\nu/2\sigma_\nu$ and its derivative $\dot \beta_\nu = \ddot{\sigma}_\nu/2\sigma_\nu - \dot\sigma_\nu^2/2\sigma_\nu^2$
yields
\begin{align}
    \frac{N}{4}\ddot\sigma_\nu = -\frac{\partial V_{\rm eff}(\sigma_\nu,\sigma_{\bar\nu})}{\partial \sigma_\nu}\label{EL_width_eff_pot}. 
\end{align}
Here $V_{\rm eff}$ is the effective potential for the oscillations of widths $\sigma_\nu$ and $N/4$ is the effective mass.
% \begin{align}
%     V_{\rm eff} = \int d\sigma_1 \frac{\partial V_{\rm eff}(\sigma_1,\sigma_2)}{\partial\sigma_1} + \int d\sigma_2 \frac{\partial V_{\rm eff}(\sigma_1,\sigma_2)}{\partial\sigma_2}\Bigg|_{\text{only terms with} \sigma_2} = \frac{4}{N}E(\sigma_1,\sigma_2)
% \end{align}
% where $E$ is the energy at equilibrium.
Linearization of Eq.~(\ref{EL_width_eff_pot}) about equilibrium widths gives
\begin{align}
    \delta \ddot \sigma_\nu = -\frac{4}{N}\frac{\partial^2 V_{\rm eff}}{\partial\sigma_\nu^2}\Bigg|_{\sigma_1=\sigma_2=\sigma} \delta\sigma_\nu -\frac{4}{N}\frac{\partial^2 V_{\rm eff}}{\partial\sigma_\nu, \partial\sigma_{\bar\nu}}\Bigg|_{\sigma_1=\sigma_2=\sigma}\delta\sigma_{\bar{\nu}}
\end{align}
where 
\begin{align}
    \omega_1^2&=\frac{4}{N}\frac{\partial^2V_{\rm eff}}{\partial \sigma_\nu^2}\Bigg|_{\sigma_1=\sigma_2=\sigma},\
    \omega_2^2=\frac{4}{N}\frac{\partial^2 V_{\rm eff}}{\partial \sigma_{\bar \nu}\partial \sigma_\nu}\Bigg|_{\sigma_1=\sigma_2=\sigma}.
\end{align}
Therefore $\omega_1$ and $\omega_2$ are entirely determined by the second derivatives of the effective potential evaluated at the equilibrium point $\epsilon(t) = 0$. These second-order derivatives of any higher-order terms in $\epsilon(t)$ (beyond the second order) in the effective potential identically evaluate to zero and do not contribute to the density- and spin-breathing-mode frequencies.

\end{widetext}
\bibliography{ref}{}
\bibliographystyle{apsrev4-1}

\end{document}